\documentclass[12pt]{article}
\usepackage[a4paper]{geometry}
\usepackage{amsmath}
\usepackage{amsthm}
\usepackage{placeins}
\usepackage[colorlinks,citecolor=blue,urlcolor=blue]{hyperref}
\usepackage[utf8]{inputenc}
\usepackage{setspace}
\usepackage{rotating}
\usepackage{color}
\usepackage{amsmath}
\usepackage{pdflscape}
\usepackage{float}

\usepackage{natbib}
\definecolor{red}{rgb}{1,0,0}
\definecolor{blue}{rgb}{0,0,1}
\definecolor{green}{rgb}{0,0.6,0.4}

\DeclareMathOperator{\var}{var}

\title{Unveiling Challenges in Mendelian Randomization for Gene-Environment Interaction}

\author{Malka Gorfine$^1$, Conghui Qu$^2$, Ulrike Peters$^2$, Li Hsu$^2$ \\
$^1$ Department of Statistics and Operations Research, Tel Aviv University, \\
Tel Aviv, Israel \\
$^2$ Division of Public Health Sciences, Fred Hutchinson Cancer Center, \\
Seattle, USA}
\begin{document}

\maketitle


\begin{abstract} 

Many diseases and traits involve a complex interplay between genes and environment, generating significant interest in studying gene-environment interaction through observational data. However, for lifestyle and environmental risk factors, they are often susceptible to unmeasured confounding factors and as a result, may bias the assessment of the joint effect of gene and environment. Recently, Mendelian randomization (MR) has evolved into a versatile method for assessing causal relationships based on observational data to account for unmeasured confounders. This approach utilizes genetic variants as instrumental variables (IVs) and aims to offer a reliable statistical test and estimation of causal effects. MR has gained substantial popularity in recent years largely due to the success of large-scale genome-wide association studies in identifying genetic variants associated with lifestyle and environmental factors. Many methods have been developed for MR; however, little work has been done for evaluating gene-environment interaction. In this paper, we focus on two primary IV approaches: the 2-stage predictor substitution (2SPS) and the 2-stage residual inclusion (2SRI), and extend them to accommodate gene-environment interaction under both the linear and logistic regression models for the continuous and binary outcomes, respectively. Extensive simulation and analytical derivations show that finding solutions in the linear regression model setting is relatively straightforward; however, the logistic regression model is significantly more complex and demands additional effort. 

\end{abstract}


\newpage

\doublespace    

\section{Introduction}
There is a great interest to study the interaction between genes and environmental risk factors in complex diseases \citep{virolainen2023gene}. However, environmental risk factors are often susceptible to unmeasured confounding, which, if not properly accounted, can yield under- or over-estimation of the effect. To study gene-environment interaction, it is inevitable to study the joint effect of gene and environment. If these factors, especially environmental factors, are subject to confounding, then proper inference of gene-environment interaction may be compromised. 

In epidemiology, the approach of Mendelian randomization (MR) has evolved into a versatile method for evaluating causal associations based on observational data, when unmeasured confounding variables are abundant. 
The approach uses genetic variants as instrumental variables (IVs),  and the goal is to provide a reliable statistical test and estimation of causal effects when unmeasured confounding factors are present. MR capitalizes on a natural experiment in which the genotypes are randomly assigned during meiosis, given the parents' genes, and it is assumed that genotypes are indirectly impact the disease status, independently from any potential confounders. A partial list of econometrics, causal and epidemiological works presenting the basic ideas and important discussions includes \cite{bowden1984instrumental,angrist1996identification,davey2003mendelian,katan2004commentary,pearl2000causality,greenland2000introduction,martens2006instrumental,hernan2006instruments} and \cite{didelez2007mendelian}.

Consider the variables $X$ and $Y$, where $X$ is the non-randomized exposure or treatment and $Y$ is the response. Intervening on $X$ refers to the act of setting $X$ to a specific value of our choice, which does not alter the distributions of the other variables in the system except through the effects induced by the changes made to $X$. Also, consider a variable denoted as $G_{IV}$, which we aim to utilize as the IV. In our context $G_{IV}$ is the genotype or a polygenic risk score (PRS). An unobservable variable $U$ represents the potential confounding that may exist between $X$ and $Y$. The following are the required assumptions related to $G_{IV}$ which define an IV:
\begin{description}
    \item[A.1] IV is associated with the exposure of interest, i.e., $G_{IV} \not\!\perp\!\!\!\perp X$.
    \item[A.2] IV is independent of unmeasured confounders of the exposure and outcome, i.e., $G_{IV} \perp\!\!\!\perp U$.
    \item[A.3] IV is independent of the outcome outside of the mediating effects of the exposure, i.e., 
    $Y \perp\!\!\!\perp G_{IV} |(X,U) $.
\end{description}

A wealth of literature is available  reviewing various aspects of MR and IVs, such as the underlying MR assumptions \citep{hernan2006instruments,didelez2007mendelian,glymour2012credible,didelez2010assumptions}, the available methods \citep{angrist1996identification,baiocchi2014instrumental}, multiple IVs \citep{palmer2012using,clarke2012instrumental}, and non-linear regression models such as logistic, Poisson, Cox proportional hazards and additive hazards  \citep{palmer2011instrumental,burgess2017review,wan2018general}. In short, two main IV approaches have been extensively developed within the context of linear models. The first approach, known as 2-stage predictor substitution (2SPS) or 2-stage least squares, starts with a  linear regression model of the exposure on the IV. The fitted value obtained from this first-stage regression is replacing the exposure in the second-stage linear regression of the outcome $Y$.  The second approach, equally popular, is called 2-stage residual inclusion (2SRI). In this approach, the residuals from the first-stage regression are included as an additional covariate, along with the exposure $X$, in the second-stage linear regression of the outcome $Y$. Both estimators, 2SPS and 2SRI, are consistent when applied to linear models \citep{terza2008two,cai2011two}. Extending 2SPS and 2SRI to address non-linear models simply involves replacing the second-stage linear regression model with alternative models such as Poisson, logistic, or Cox proportional hazard models.

As nicely summarized by \cite{wan2018general}, the literature has conflicting views on the consistency of two-stage IV methods for nonlinear models. Focusing on the conditional treatment effect based on observational data, \cite{terza2008two} showed the consistency of 2SRI in a wide range of nonlinear models and under restrictive assumptions, while this consistency has not been established for 2SPS. As a consequence, 2SRI has gained widespread acceptance as the preferred method in studies that deal with discrete and survival outcomes \citep{wan2018general}. Conversely, within the context of MR, both the 2SRI and 2SPS approaches have demonstrated bias when estimating the log odds ratio, where a dichotomous genetic marker was utilized as IV and the exposure $X$ was a continuous phenotype. Studies have revealed that this bias amplifies with a greater magnitude of unmeasured confounding \citep{palmer2008adjusting,burgess2013identifying,wan2018general}. Methods leveraging gene-environment interactions within MR analyses \citep{spiller2019detecting,tchetgen2021genius,spiller2022interaction} estimate causal associations while correcting for instrumental invalidity, in particular when assumptions A.2 or A.3 are violated.

Because of conflicting recommendations in the existing literature and because of the increasing popularity of applying
2SRI to control for unmeasured confounding in clinical studies, \cite{wan2018general}  revisited  the conditions under which one can establish the consistency of 2-stage IV nonlinear models. Their main conclusions include:
\begin{enumerate}
    \item Previous findings on 2SRI \citep{terza2008two}  rely on an (unrealistic) underlying assumption that only one unmeasured confounder exists.
    \item When estimating non-null conditional treatment effect, 2SRI suffers from the same problem as 2SPS does and is not consistent in non-collapsible models even given a perfect IV.
    \item 2SRI and 2SPS estimators are consistent when
estimating collapsible effect measures (such as risk rates or risk differences) using Poisson or additive hazards models.
    \item In practice, one may consider using collapsible models instead of non-collapsible ones. For example, in case of rare binary events, one may consider log-linear models instead of logistic models, if possible. 
\end{enumerate}

\cite{terza2008two} and \cite{burgess2017review} also considered an adjusted 2SPS method where the fitted value of exposure $X$ and the residual from the first-stage regression are included additively in the second-stage regression of $Y$. Some investigators have therefore recommended this adjusted two-stage method when the second-stage regression is logistic on the premise that it is less biased than the unadjusted two-stage method \citep{burgess2017review}. The discussion so far assumed no interaction between the exposure $X$ and any other observed variables, here, genes, in the regression model of $Y$.

The present study focuses on testing and estimating the effect of interaction between $G$ and $X$ on $Y$, as well as the main effects of $G$ and $X$, which are necessary to facilitate the inference of the joint effect size of $G$ and $X$. Here, $Y$ may either be a continuous or a dichotomous variable, and linear and logistic regression models are adopted, respectively. For each model we explore the naive estimators that ignore the existence of unmeasured confounders, and also straightforward modified versions of 2SPS and 2SRI to accommodate effect modification. By extensive simulation the biases of causal effects and type-I error of these estimators are demonstrated. 

Our work offers three key contributions: (i) We demonstrate the validity of straightforward extensions of MR methods designed for linear outcomes $Y$ when dealing with gene-environment interaction. These extensions are valid in terms of bias and type-I error rates. (ii) In the context of binary $Y$ and logistic regression, we elucidate the significant distinctions between the 2SPS and 2SRI methods. Furthermore, we reveal that none of these methods exhibit consistency; however, they generally maintained type I error for gene-environment interaction or with limited inflation, except for some scenarios. The quest for a valid approach in the realm of non-linear regression models remains an unresolved challenge. (iii) We apply the MR methods to a large-scale gene-environment interaction analysis of BMI in association with colorectal cancer risk, encompassing 44,500 cases and 52,235 controls, and compare the performance of these methods in a real data setting. While there is no evidence of type-I error inflation for gene-BMI interaction, it remains unclear what the true effect sizes of gene-BMI interaction should be from the MR results. Interestingly, we demonstrate that the true causal effect size of BMI is likely greater than the effect size based on measured BMI.  

\section{Linear Regression}
We start with a continuous outcome $Y$ such that the true model is given by
\begin{equation}
    Y  =  \beta_0 + \beta_1 X + \beta_2 G + \beta_3 X G + \beta_Z Z + \beta_U U + \epsilon_Y,  \label{eq:y1cont}
    \end{equation}
where $U$ represents unmeasured confounding variable(s) that is also affecting $X$, $Z$ is an observed covariate affecting $Y$ and $X$, $\beta_j, j = 1, \ldots, 3,$  and $\beta_Z$ are regression coefficients, and $\epsilon_Y$ is the residual error with mean 0 and variance $\sigma_Y^2$, and  is independent of $(X,G,Z,U)$. For simplicity of presentation, a univariate $Z$ is considered.  Without loss of generality, it is assumed that $U$ is a mean 0 and variance $\sigma_U^2$ random variable and is independent of $Z$. By the MR assumption, $U$ is independent of $G$ as well as the genetic instruments defined below for the model of the exposure $X$. The coefficient $\beta_3$ represents the effect-modification parameter. Our goal is estimating $\beta=(\beta_1,\beta_2,\beta_3)^T$ and testing the null hypothesis $H_0: \beta_3=0$ against a two-sided alternative.


In this work we focus on the setting in which there are an observed confounder $Z$ and an unobserved confounder $U$. In addition, it is assumed genetic instrument variable $G_{IV}$ is available and has a relationship with $X$ as described in the following
\begin{equation}
    X  =   \gamma_0 + \gamma_{IV} G_{IV} +  \gamma_Z Z + \gamma_U U + \epsilon_X, \label{eq:x1cont} 
\end{equation}
 where $\gamma_0, \gamma_{IV}$, $\gamma_Z$, and $\gamma_U$ are the regression coefficients,  $\epsilon_X$ is the residual error with mean 0 and variance 1 and is independent of $(G_{IV},Z,U)$ and $Y$.

 As $U$ is unobserved, modeling $X$ relies on the observable data by fitting a linear regression model, namely, 
 $$
 E(X|G_{IV},Z) = \alpha_0 + \alpha_1 G_{IV} + \alpha_Z Z \, 
 $$ 
 and getting $(\widehat{\alpha}_0,\widehat{\alpha}_1, \widehat{\alpha}_Z)$ by using e.g., ordinary least squares. Given that $U$ and $\epsilon_X$ are independent, the estimator $(\widehat{\alpha}_0,\widehat{\alpha}_1, \widehat{\alpha}_Z)$ is consistent to the true value of $(\gamma_0, \gamma_{IV}, \gamma_Z)$ in model (\ref{eq:x1cont}) and has the usual asymptotic normality property following the Central Limit Theorem. 
 Then, $X$ can be predicted by 
 $$\widehat X = \widehat \alpha_0 + \widehat \alpha_1 G_{IV} + \widehat \alpha_Z Z $$ 
 and the residuals are given by 
 $$
 \widehat \delta = X - \widehat X \, .
 $$ 
 Alternatively, we can consider $X$ to be predicted only by $G_{IV}$, i.e., 
 $$
 \widehat X_a = \widehat \alpha_1 G_{IV} \, ,
 $$
so the residual is given by
$$
\widehat \delta_a = X - \widehat X_a \, .
$$ 
 
In the subsequent discussion, we consider six methods to estimate $\beta$, denoted as $\widehat{\theta} = (\widehat \theta_1, \widehat \theta_2, \widehat \theta_3)^T$. It is important to note that each approach computes $\widehat{\theta}$ differently, although we use a unified notation for ease of presentation:
 \begin{description}
 \item[1. Naive:] The naive estimator  ignores the fact that there could be unobserved confounders and uses the following model  $E(Y|X,G,Z)  =  \theta_0 + \theta_1  X + \theta_2 G + \theta_3 X G + \theta_Z Z$ to get  $\widehat{\theta}$.
\item[2. 2SPS:] The two-stage prediction substitution (2SPS) under Eq.~(\ref{eq:y1cont}) starts with the first-stage regression, where the exposure $X$ is regressed on the IVs, $G_{IV}$, and known confounders $Z$  to obtain fitted values of the exposure, $\widehat{X}$. In the second-stage regression, the outcome $Y$ is regressed on the fitted values
for the exposure $\widehat{X}$ from the first-stage regression, i.e., 
 $E(Y|\widehat{X},G,Z)  =  \theta_0 + \theta_1 \widehat X + \theta_2 G + \theta_3 \widehat X G + \theta_Z Z$.
 \item[3. 2SPSadj:] Based on the regression model of $X$,  the residual, $\widehat{\delta} = X-\widehat X$, is obtained, which comprises of both the unobserved confounders $U$ and the  error $\epsilon_X$. In the second-stage regression,   the outcome $Y$ is regressed not only on the fitted values
for the exposure $\widehat{X}$ and known confounders $Z$, but also $\widehat{\delta}$ in order to capture some of the unobserved confounders $U$. That is, the second-stage model is given by
 $E(Y|\widehat{X},G,Z,\widehat{\delta})  =  \theta_0 + \theta_1 \widehat X + \theta_2 G + \theta_3 \widehat X G + \theta_Z Z + \theta_4 \widehat{\delta}$, with further adjustment of $\widehat{\delta}$.
 \item[4. 2SPSa:] Sometimes only the regression coefficient associated with $G_{IV}$ is available. An alternative to the 2SPS method is to obtain the fitted value for $X$ based  only on $G_{IV}$, i.e., $\widehat X_a = \widehat \alpha_1 G_{IV}$. In the second-stage regression for $Y$, the incorporation of the interaction terms between $Z$ and $G$ is required in order to absorb the effect of $Z$ on $X$. Hence, the outcome regression model is given by
 $E(Y|\widehat{X}_a,G,Z)  =  \theta_0 + \theta_1 \widehat{X}_a + \theta_2 G + \theta_3 \widehat{X}_a G + \theta_Z Z + \theta_{GZ} Z G$.
 \item[5. 2SPSadj-a:] The second stage regression will additionally include the residuals $\delta_a = X - \widehat X_a$ similarly to the adjusted 2SPS estimation method, namely,
$E(Y|\widehat{X}_a,G,Z,\widehat{\delta}_a)  =  \theta_0 + \theta_1 \widehat{X}_a + \theta_2 G + \theta_3 \widehat{X}_a G + \theta_Z Z + \theta_{GZ} Z G + \theta_4 \widehat{\delta}_a$. 
\item[6. 2SRI:] The two-stage residual
inclusion (2SRI) under Eq.~(\ref{eq:y1cont}) is an adjusted two-stage method, where the residual $\widehat{\delta}$ from the first-stage regression (\ref{eq:x1cont}) is included
in the second-stage regression (i.e., Eq.~(\ref{eq:y1cont})) \citep[and references therein]{terza2008two}. The second-stage model is given by $E(Y|X,G,Z,\widehat{\delta}) = \theta_0 + \theta_1  X + \theta_2 G + \theta_3 X G + \theta_Z Z + \theta_4 \widehat\delta$.
 \end{description}

\subsection{Justification of 2SPS and 2SRI approaches}

We begin by taking a closer look at the 2SPS approach. Plugging in Eq.~(\ref{eq:x1cont}) into Eq.~(\ref{eq:y1cont}), we obtain
\begin{eqnarray*}
    Y & = & \beta_0 + \beta_1 ( \gamma_0 + \gamma_{IV} G_{IV} +  \gamma_Z Z + \gamma_U U + \epsilon_X) + \beta_2 G \\
    & & + \beta_3 ( \gamma_0 + \gamma_{IV} G_{IV} +  \gamma_Z Z + \gamma_U U  + \epsilon_X) G + \beta_Z Z + \beta_U U + \epsilon_Y, \\
    & = & {\beta}_0 + \beta_1 \widetilde X + (\beta_2 +  \beta_3 \gamma_U U  +  \beta_3 \epsilon_X)  G  + \beta_3 \widetilde X G  \\
    & & + \beta_Z Z + (\beta_1 \gamma_U U  + \beta_U U) + \beta_1 \epsilon_X + \epsilon_Y
\end{eqnarray*}
where
$$
\widetilde X = \gamma_0 + \gamma_{IV} G_{IV} +  \gamma_Z Z \, .
$$
Then, taking the conditional expectation of $Y$  given $G$, $G_{IV}$, and $Z$, we have
\begin{eqnarray*}
    E(Y|G, G_{IV}, Z) & = & {\beta}_0 + \beta_1 \widetilde X + \beta_2  G + \beta_3 \widetilde X G + \beta_Z Z \, .  
\end{eqnarray*}
This supports the use of the 2SPS approach, wherein fitting a model of $Y$ with $(\widehat X,G,\widehat X G,Z)$  results in a consistent estimator for $\beta=(\beta_1,\beta_2,\beta_3)$. This justification stems from the collapsibility property of the linear regression model, allowing for the substitution of $\widetilde X$ with $\widehat X$.
Furthermore, including $\widehat{\delta}$ in the model for the adjusted 2SPS approach does not compromise the consistency of the estimators for $\beta$; however, it has the potential to decrease the standard errors.
This is due to the assumption that $\widehat{\delta}$ is independent of $\widehat X$ and encompasses the effects of $U$. If the influence of $U$ is large, substantial gain in efficiency can be achieved without jeopardizing the consistency of the estimators.

Next, when we contemplate an alternative predictor  $\widetilde X_a =  \gamma_{IV} G_{IV}$, it yields
\begin{eqnarray*}
    Y & = & \beta_0 + \beta_1 ( \gamma_0 + \widetilde X_a +  \gamma_Z Z + \gamma_U U  + \epsilon_X) + \beta_2 G \\
    & & + \beta_3 ( \gamma_0 + \widetilde X_a +  \gamma_Z Z + \gamma_U U  + \epsilon_X) G + \beta_Z Z + \beta_U U + \epsilon_Y, \\
    & = & \widetilde{\beta}_0 + \beta_1 \widetilde X_a + (\beta_2 + \beta_3 \gamma_0 +  \beta_3 \gamma_Z Z + \beta_3 \gamma_U U  +  \beta_3 \epsilon_X)  G  + \beta_3 \widetilde X_a G  \\
    & & + \widetilde{\beta}_Z Z + (\beta_1 \gamma_U U  + \beta_U U) + \beta_1 \epsilon_X + \epsilon_Y,
\end{eqnarray*}
where $\widetilde{\beta}_0$ and $\widetilde{\beta}_Z$ are generic symbols for the intercept and the regression coefficient associated with $Z$. The expectation of $Y$ conditional on $G$, $G_{IV}$ and $Z$, gives
\begin{eqnarray*}
    E(Y|G, G_{IV}, Z) & = & \widetilde{\beta}_0 + \beta_1 \widetilde X_a + (\beta_2 + \beta_3 \gamma_0) G + \beta_3 \widetilde X_a G + \widetilde{\beta}_Z Z + \beta_{GZ} G Z \, ,
\end{eqnarray*}
where $\beta_{GZ}$ is the regression coefficent for the interaction term between $G$ and $Z$. Drawing from the expectation outlined above, modeling $Y$ with $(G,\widehat X_a, G \widehat X_a, Z, GZ)$ produces estimates that are consistent for $\beta_1$ and the interaction term $\beta_3$. However, it does not yield consistent estimates for the main effect of $G$, except for when $\gamma_0 = 0$. Notably, in contrast to the adjusted 2SPS approach, this method necessitates the inclusion of an additional interaction term $GZ$.

Moving on, let us delve into the 2SRI approach. We make an assumption that the joint distribution of $(X,U)$, given $G_{IV}$ and $Z$, conforms to a bivariate normal distribution with 
\begin{eqnarray*}
\left (
    \begin{array}{c}
    U \\
    X
    \end{array}
    \right ) | \, G_{IV},Z & \sim & N \left( \left ( \begin{array}{c}
    0 \\
    \mu_1 
    \end{array} \right ), \left (
\begin{array}{cc}
\sigma_U^2 & \gamma_U \sigma_U^2 \\
\gamma_U \sigma_U^2 & \sigma_X^2 
\end{array}
    \right )
    \right)
\end{eqnarray*}
where $\mu_1 = \gamma_0 + \gamma_{IV} G_{IV} + \gamma_Z Z$ and $\sigma_X^2 = \gamma_U^2 \sigma_U^2 + \sigma_{\epsilon_X}^2$. Hence, the conditional distribution of $U$ given $X$ is 
\begin{eqnarray*}
    U|X \sim N \left ( \gamma_U \sigma_U^2/\sigma_X^2(X-\mu_1), \sigma_U^2 ( 1- \gamma_U^2\sigma_U^2/\sigma_X^2 ) \right) \, .
\end{eqnarray*}
Lastly, in the context of Eq.~(\ref{eq:y1cont}), we can calculate the expected value of $Y$ given $(X,G,Z)$, resulting in
\begin{eqnarray*}
E(Y|X, G, Z) & = & \beta_0 + \beta_1 X + \beta_2 G + \beta_3 X G + \beta_U E(U|X, G) + \beta_Z Z \\
& = & \beta_0 + \beta_1 X + \beta_2 G + \beta_3 X G + \beta_U \frac{\gamma_U \sigma_U^2}{\sigma_X^2}(X-\mu_1) + \beta_Z Z
\end{eqnarray*}
and $\delta = X - \mu_1$. This computation provides a rationale for utilizing the 2SRI estimator in the linear model. This also suggests that the naive method provides a consistent estimator for the main effect of $G$, $\beta_2$, and the interaction effect $\beta_3$, while it does not yield a consistent estimator for the main effect $X$, $\beta_1$, the latter of which has also been proposed by \cite{vanderweele2012sensitivity}. It is worth noting that even if we substitute $\mu_1$ with $\gamma_{IV} G_{IV}$, the 2SRI estimator remains consistent for $\beta$, as the influence of $Z$ is absorbed into the main effect of $Z$.

\subsection{Simulation Study - Linear Regression}

A total of $10,000$ observations were considered for Model (\ref{eq:x1cont}) for the exposure $X$ with $\gamma_0=0$, $\gamma_Z=\gamma_{IV}=0.5$,  $G_{IV} \sim \mbox{N}(0,1)$, $U \sim \mbox{N}(0,1)$, $Z \sim \mbox{N}(0,1)$ and $\epsilon_X \sim N(0,1)$. The instrumental variable $G_{IV}$ represents a polygenic risk score. Multiple values of $\gamma_U$ were explored, $\gamma_U \in \{0,0.5,1,2,4\}$. The values of $X$  underwent a transformation to standardize their variance to 1 before being used in the primary $Y$ outcome model to facilitate ease of comparison.  
For the linear regression model of Eq.~(\ref{eq:y1cont}) for the outcome $Y$, $10,000$ observations were generated with $G \sim \mbox{Bin}(2,0.3)$ representing the number of copies (0, 1 or 2) of a single bi-allelic SNP. Also,
    $(\beta_0,\beta_1,\beta_2,\beta_Z)=(0,1,0.5,0.5)$, multiple values of $\beta_U$, $\beta_U \in \{0,1.5,3\}$ and $\epsilon_Y \sim N(0,1)$. 
   Each configuration was examined twice: once with $\beta_3=0.5$ and once under the null hypothesis, $H_0: \beta_3=0$, to explore whether the presence of the unmeasured variable $U$ has an impact on the type-I error. Unless stated otherwise, the simulation results are derived from 500 repetitions of each configuration.
   We also included a scenario in which $X$ and $G$ are dependent through a dependence between $G_{IV}$ and $G$. In particular, the following conditional distribution is adopted 
    $$
    G|G_{IV} \sim \mbox{Bin}(2 \, , \, 0.3+0.3I\{G_{IV}>0\}) \, . 
    $$  
    Table \ref{tbl:sim-desc} provides a summary of the configurations studied.

\begin{figure}[H]
\centering
\includegraphics[width=14cm,height=10cm]{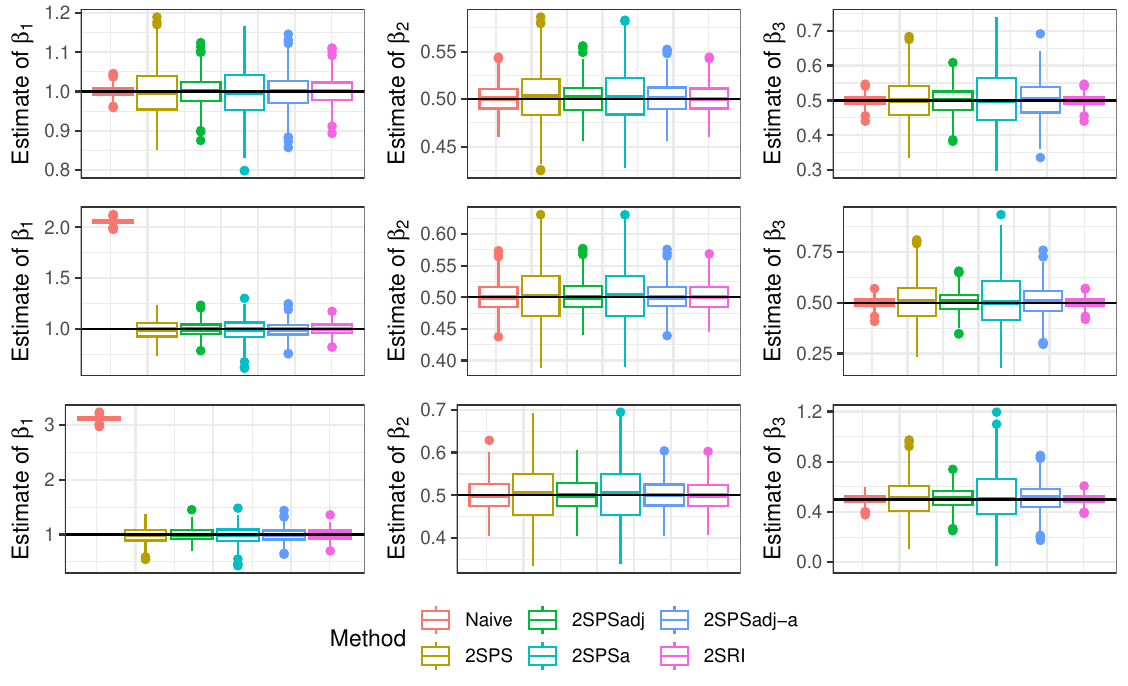}\\
\vspace{-0.05cm} 
\includegraphics[width=14cm,height=10cm]{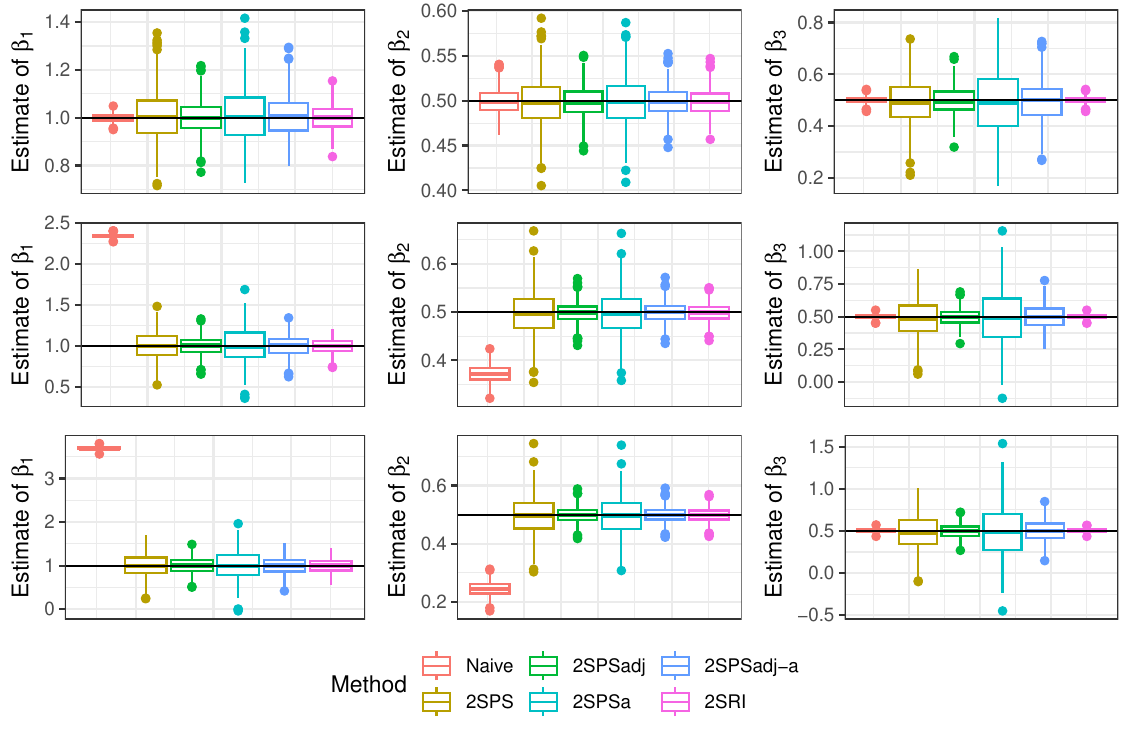}
\caption{Simulation results, linear regression, Setting I.A with $\gamma_U=1$ (top sub-figure) and Setting II.A with $\gamma_U=1$ (bottom sub-figure): within each sub-figure,  top line $\beta_U=0$, middle line $\beta_U=1.5$ and third line $\beta_U=3$.}\label{fig:linear1A}
\end{figure}

\begin{figure}[H]
\centering
\includegraphics[width=14cm,height=10cm]{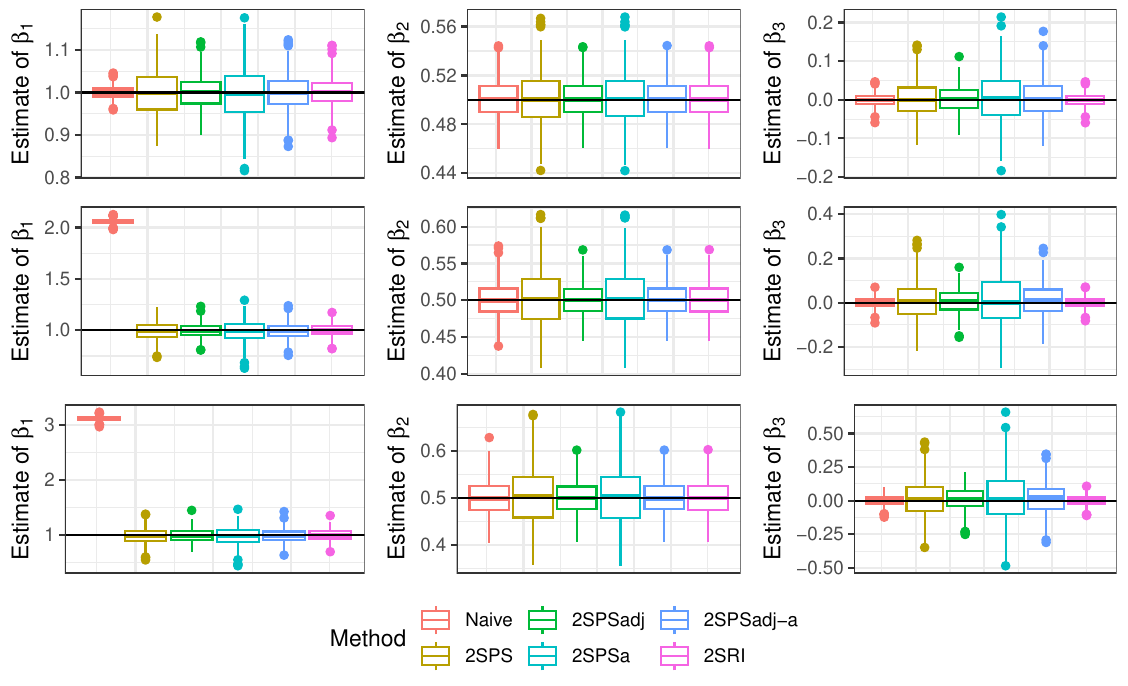}\\
\vspace{-0.05cm} 
\includegraphics[width=14cm,height=10cm]{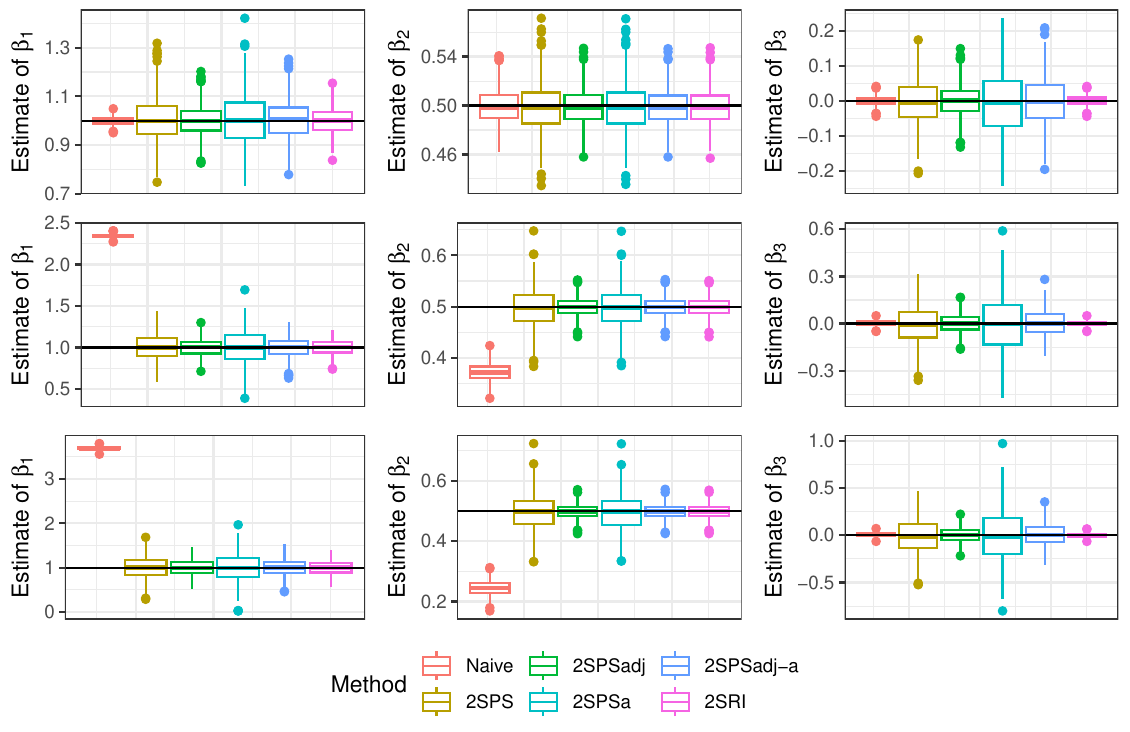}
\caption{Simulation results, linear regression, Setting I.B with $\gamma_U=1$, $\beta_3=0$ (top sub-figure) and Setting II.B with $\gamma_U=1$, $\beta_3=0$ (bottom sub-figure): within each sub-figure,  top line $\beta_U=0$, middle line $\beta_U=1.5$ and third line $\beta_U=3$.}\label{fig:linear2B}
\end{figure}

Figures \ref{fig:linear1A}-\ref{fig:linear2B}, Tables \ref{tbl:GXIblinear}-\ref{tbl:GXIIblinear} and Tables S1-S4 of the Supporting Information (SI) summarise the results. Figures \ref{fig:linear1A}-\ref{fig:linear2B} display results specifically for  $\gamma_U=1$ as the overall findings remain consistent across various values of $\gamma_U$. The complete results are in Tables S1-S4 of SI.
Biased results are marked in bold.  

As expected, when $\beta_U=0$, all the methods yield unbiased estimators for $\beta=(\beta_1,\beta_2,\beta_3)$ across all scenarios. Notably, in many cases, the naive approach demonstrates significantly higher efficiency compared to the other five estimators.  In cases where $\beta_U > 0$ and there is independence between $G$ and $G_{IV}$, the naive approach exhibits bias in estimating $\beta_1$ while maintaining its unbiasedness in estimating $(\beta_2,\beta_3)$. All MR-based estimators are unbiased for $\beta$. When $G$ and $G_{IV}$ are dependent, the naive estimator exhibits bias in estimating $(\beta_1,\beta_2)$, while the MR estimators provide unbiased estimations for $\beta$. Among the five MR-based estimators, 2SRI demonstrates the highest level of efficiency, comparable to the naive estimator. For the 2SPS estimators, adjusting for $\widehat \delta$ improves the efficiency and 2SPS estimators that include both genetic instrument and non-genetic predictors have better efficiency. The type-I error rates of the naive and the MR-based approaches for testing $H_0:\beta_3=0$ against a two-sided alternative are reasonably close to the nominal level, 0.05. 


In conclusion, all methods, including the naive approach, are suitable for conducting an interaction test. However, the naive estimator yields biased estimators for main effects, especially the main effect of the exposure, which makes it difficult to gauge the joint effect of exposure and $G$ accurately. Each of the five MR-based techniques demonstrates consistency in estimating both the main effects and interaction effect. However, among them, 2SRI emerges as the favored option owing to its superior efficiency.

\section{Logistic Regression}
We are interested in assessing the association of exposure $X$, genotype $G$ and their interaction with binary outcome $Y$. Define $\text{logit}(p)=\log \{p/(1-p)\}$. Suppose the true model is 
\begin{eqnarray}
    \text{logit}\{ \Pr(Y = 1|X,G, Z, U) \} & = & \beta_0 + \beta_1 X + \beta_2 G + \beta_3 X G + \beta_Z Z + \beta_U U ,  \label{eq:ylogistic}
\end{eqnarray}
where $(X,G,Z,U)$ are  defined in Section 2, and we have a genetic instrument variable $G_{IV}$ for exposure $X$ and the model of Eq.~(\ref{eq:x1cont})
holds. Similar to the linear setting, we adapted the naive and the five MR-based approaches to fit the logistic regression model. Specifically, when performing the second-stage regression of $Y$ for each of the five estimators discussed in Section 2, the linear model should be substituted with a logistic regression model incorporating the relevant components.

While modifying the MR-based estimators to fit the logistic regression model is simple, justifying these MR-based estimators is challenging due to the to non-linear structure of the logit link. Here we  examine the analytical form for the 2SRI estimator . As $U$ is not directly observable, we focus on the expected value of $Y$ given $(X, G, Z)$ using an approximation method inspired by \cite{carroll2006measurement}. This approximation leverages the relationship between probit and logit models, particularly in the context of measurement error. Namely,
\begin{eqnarray*}
     \text{logit}\{E(Y|X, G, Z)\} & \approx & \left \{ {\beta}_0 + \beta_1 X + \beta_2 G + \beta_3 G X + \beta_U E(U|X, G, Z)  + \beta_Z Z \right \} / \phi \\
    & = & \left \{ {\beta}_0 + \beta_1 X + \beta_2 G + \beta_3 G X + \beta_U  \frac{\gamma_U \sigma_U^2}{\sigma_X^2}(X-\mu_1) + \beta_Z Z \right \} / \phi \, ,
\end{eqnarray*}
where 
$$
    \phi =  \left (1 + \beta_U^2 \var(U|X) /1.7^2 \right )^{1/2} \, ,
$$
and $\var(U|X) = \sigma_U^2 ( 1- {\gamma_U^2 \sigma_U^2}/{\sigma_X^2})$. Testing, for instance, the hypothesis that $\beta_3=0$ is approximately equivalent to testing $\beta_3/\phi_1=0$. This may also  explain that for the naive approach where $\delta = X - \mu_1$ is not included, the bias only impacts the main effect of $X$, but not the main effect of $G$ and the interaction coefficient $\beta_3$. Adding $\widehat{\delta}$, as in the 2SRI approach, corrects the bias for the main effect of $X$ to some extent. Even in this situation, all parameters are attenuated by $\left (1 + \beta_U^2 \var(U|X) /1.7^2 \right )^{1/2}$ and when $\beta_U = 0$, the 2SRI estimator is consistent. We attempted to derive a similar analytical approximation for the 2SPS estimators; however, the approximation is rather complex and does not have a close form. 

\subsection{Simulation Study - Logistic Regression}

A case-control study design with logistic regression model was employed, involving 5,000 cases and 5,000 controls, unless otherwise specified. The exposure $X$ was generated based on the regression model of Eq.~(\ref{eq:x1cont}) as described in Section 2.2. For the outcome of logistic regression  model of Eq. (\ref{eq:ylogistic}),     $(\beta_0,\beta_1,\beta_2,\beta_Z)=(\log(0.01/0.99),1,0.5,0.5)$ and  $\beta_U \in \{0,1.5,3\}$. The remaining elements adhere to the description provided in Section 2.2.

\begin{figure}[H]
    \centering
    \includegraphics[width=16cm,height=20cm]{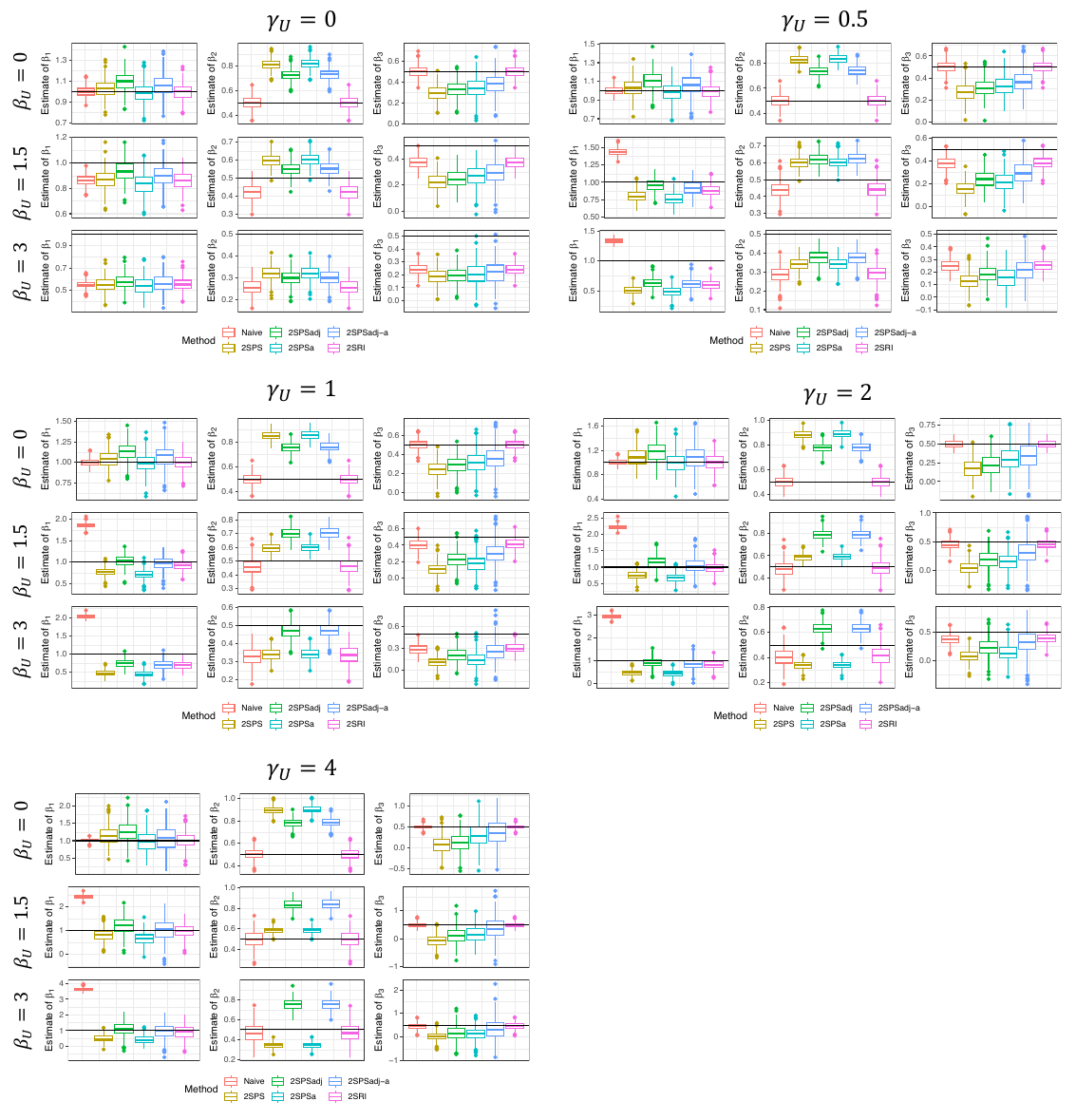}
    \caption{Simulation results, logistic regression, Setting III.A.}
    \label{fig:logistic3A}
\end{figure}

\begin{figure}[H]
    \centering
    \includegraphics[width=16cm,height=20cm]{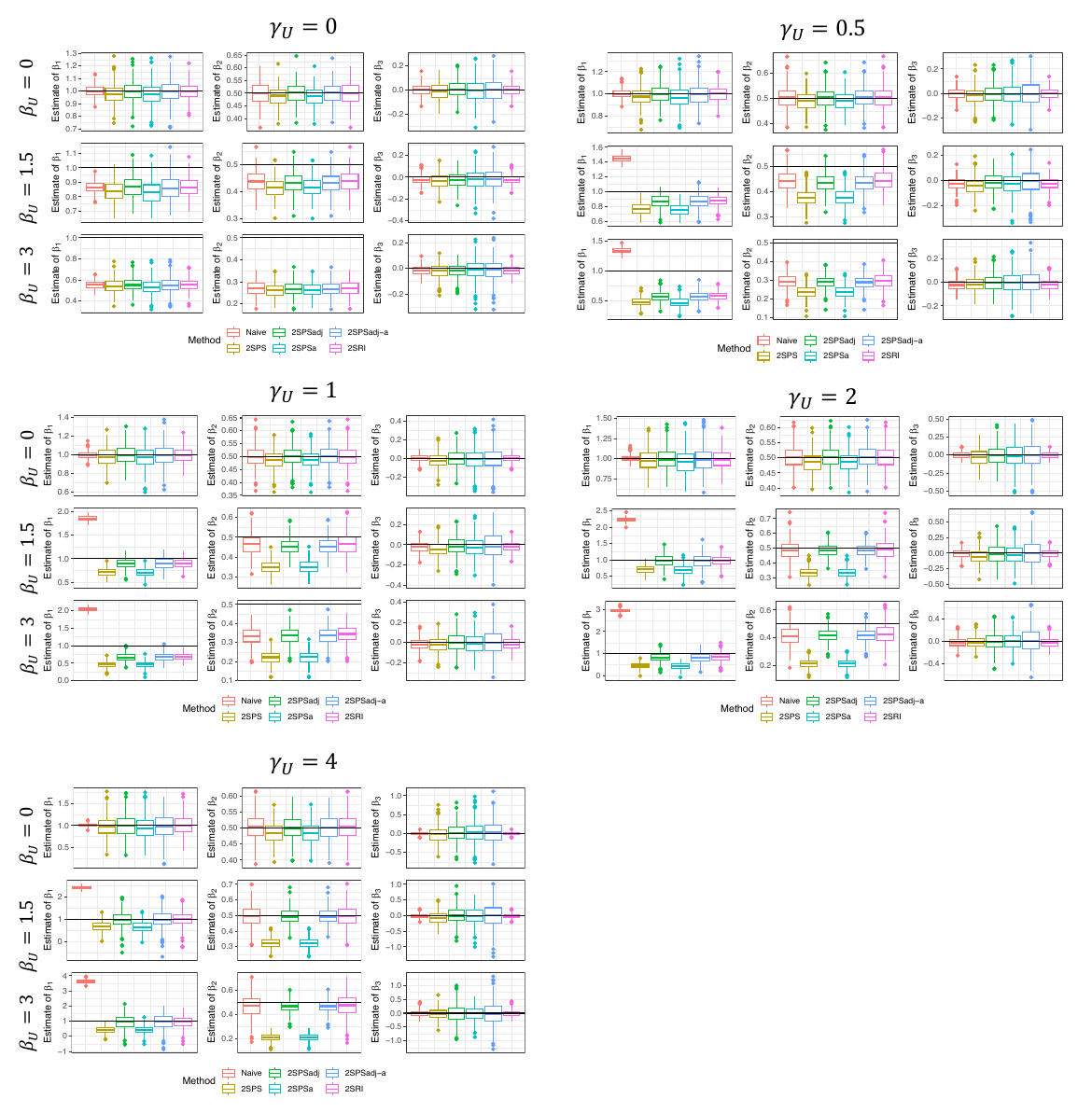}
    \caption{Simulation results, logistic regression, Setting III.B.}
    \label{fig:logistic3B}
\end{figure}

\begin{figure}[H]
    \centering
    \includegraphics[width=16cm,height=20cm]{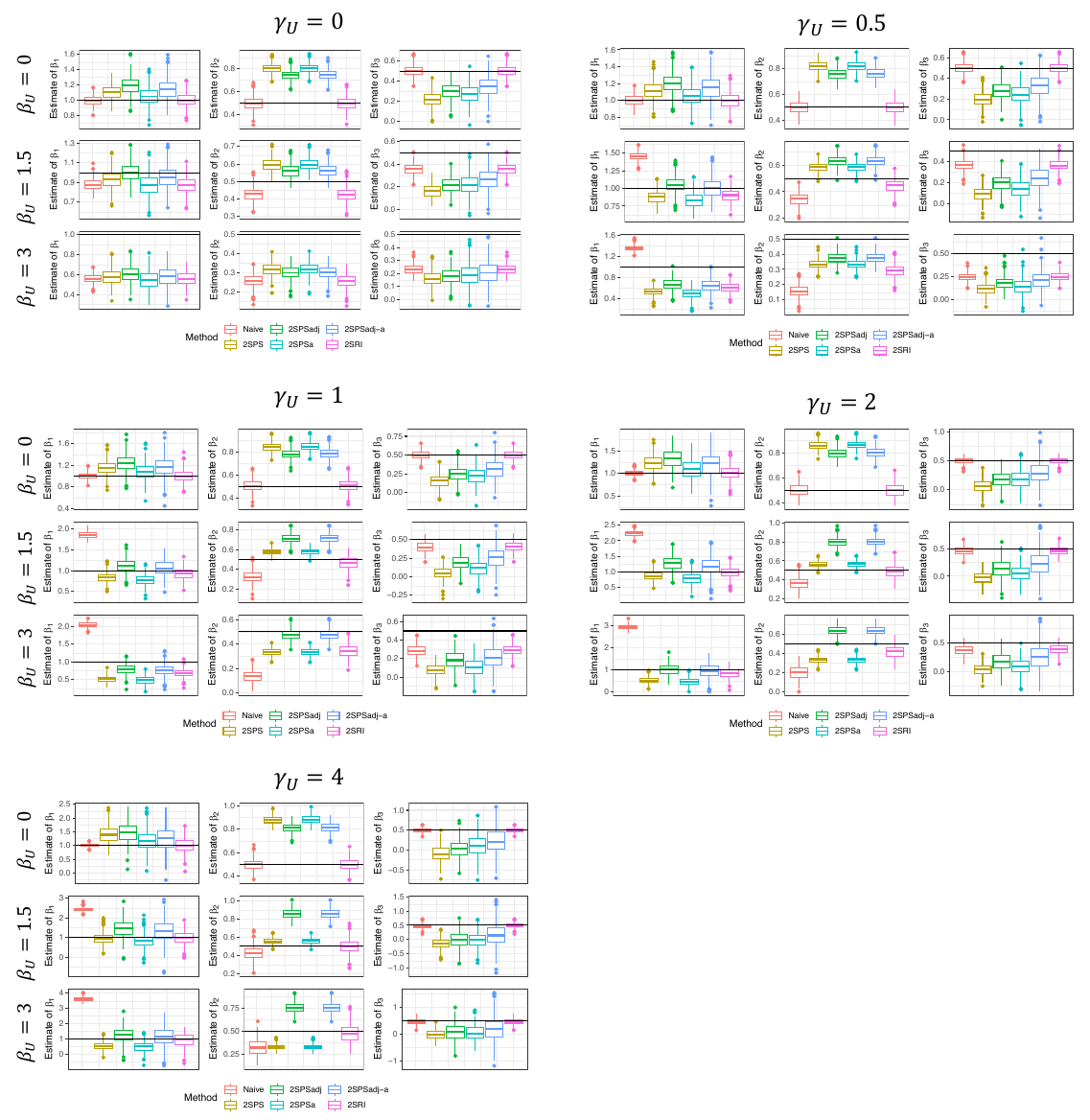}
    \caption{Simulation results, logistic regression, Setting IV.A.}
    \label{fig:logistic4A}
\end{figure}

\begin{figure}[H]
    \centering
    \includegraphics[width=16cm,height=20cm]{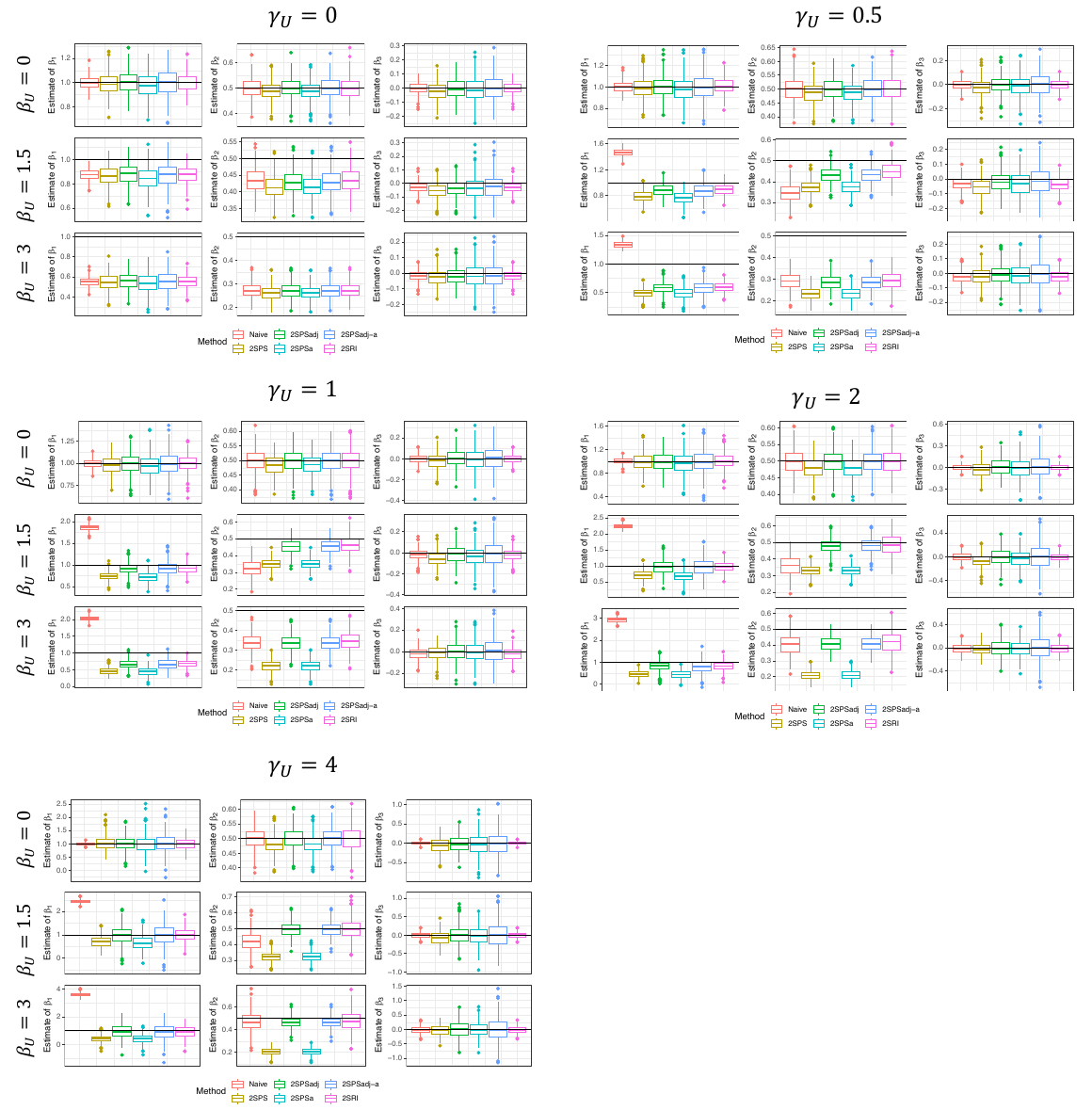}
    \caption{Simulation results, logistic regression, Setting IV.B.}
    \label{fig:logistic4B}
\end{figure}

\begin{figure}[H]
    \centering
    \includegraphics[width=16cm,height=20cm]{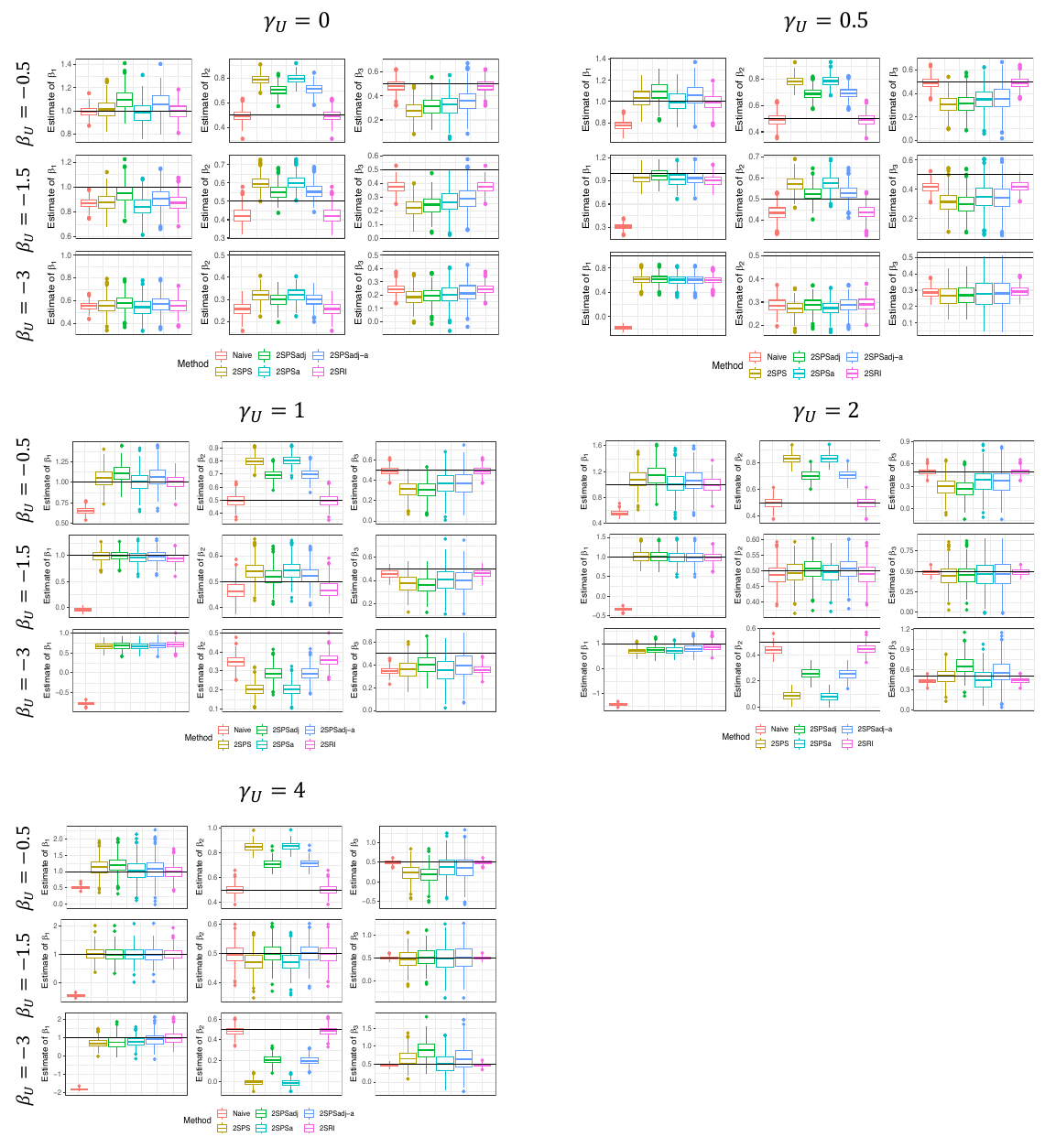}
    \caption{Simulation results, logistic regression, Setting III.A with negative values of $\beta_U$.}
    \label{fig:logistic3Anegative}
\end{figure}

\begin{figure}[H]
    \centering
    \includegraphics[width=16cm,height=20cm]{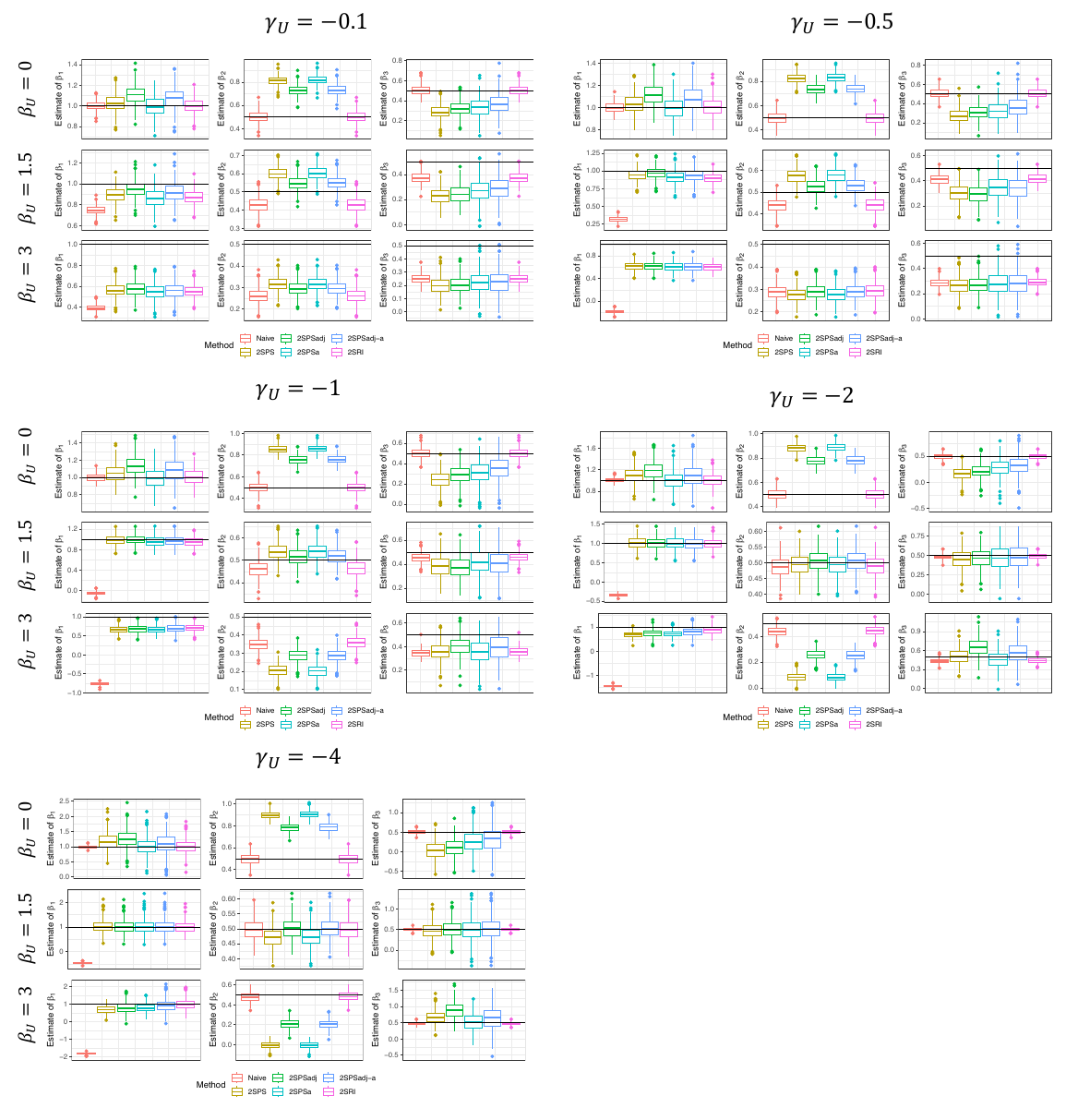}
    \caption{Simulation results, logistic regression, Setting III.A with negative values of $\gamma_U$.}
    \label{fig:logistic3Anegative2}
\end{figure}

The results are summarized in Figures \ref{fig:logistic3A}-\ref{fig:logistic3Anegative2}. 
The naive estimator for $(\beta_1, \beta_2, \beta_3)$ exhibits bias if $\beta_U \ne 0$, even when $\gamma_U=0$.  Notably, unlike the linear model, this bias pertains to each parameter, including $\beta_3$ for the effect of GxE, and is not limited to just $\beta_1$ for the effect of exposure $X$. This bias arises from the non-collapsibility property inherent to logistic regression.  While the naive estimator for $\beta_2$ and $\beta_3$ is generally biased downward, when $\gamma_U>0$ and $\beta_U = 1.5$ or $3$, the naive estimator  overestimates $\beta_1$ (Figures \ref{fig:logistic3A}-\ref{fig:logistic4A}). When $\beta_U = -3.0, -1.5$, or $-0.5$ in the opposite direction of $\gamma_U > 0$, the naive estimator  underestimates $\beta_1$ (Figure \ref{fig:logistic3Anegative}).
A similar observation is for when $\beta_U$ is positive and $\gamma_U$ is negative (Figure~\ref{fig:logistic3Anegative2}). 

A concerning aspect of the 2SPS estimators is their notable bias, even when the naive approach exhibits only a minor empirical bias for the interaction parameter $\beta_3$ (Figure~\ref{fig:logistic3A} and \ref{fig:logistic4A}). The adjusted 2SPS estimators, 2SPSadj and 2SPSadj-a, improves the bias to some extent, especially when $\gamma_U$ is large. It appears that the only potentially valuable MR-based estimator is the 2SRI, where it occasionally improves bias for $\beta_3$. For the main effects, $\beta_1$ and $\beta_2$, the 2SPS estimators again exhibit notable bias. In contrast, the 2SRI estimator that includes the additional residual term $\widehat \delta$ frequently demonstrates lower bias for $\beta_1$ compared to the naive estimator, which can be substantial, and improves bias for $\beta_2$ occasionally. The naive and 2SRI estimators display a high degree of correlation. As in the linear model of $Y$, among all the MR-based estimators, 2SRI has the smallest standard error. 


Tables \ref{tbl:logistic3B}-\ref{tbl:logistic4B} show the results of type-I error of testing the null hypothesis of $\beta_3 = 0$. The type-I error of the naive and MR-based approaches exhibit occasional inflation when $\beta_U>0$, particularly noticeable in Setting IV. Consequently, we increased the overall sample size from 10,000 to 30,000 and expanded the number of replications from 500 to 1000 and we also explored additional scenarios, summarized in Table \ref{tbl:logisticMore}. Our primary observation is that, under small values of $\beta_U$ (e.g., $\beta_U=0.5$), the naive and 2SRI estimators tend to closely approximate the nominal type-I error. However, under large values of $\beta_U$, the statistical tests of the naive and 2SRI approaches may become invalid. The 2SPS estimator sometimes have noticeable type-I error inflation when the effect for the known confounder $Z$ is nonzero (e.g., $\beta_U = 0.5, \beta_Z = \gamma_Z = 0.5, \beta_2 = 0.5$), but the adjusted 2SPS estimator reduces the type-I error inflation somewhat, except for when $\gamma_U = 0$. The 2SPS alternative estimator based on only the genetic instrument generally approximate the type-I error well with occasionally limited inflation. When the effect of $U$ on $X$, i.e., $\gamma_U$, is large, all estimators appear to have a reasonable type-I error. Interestingly, when the main effect of $G$ is null, all estimators, including the naive estimator, maintains the correct type-I error. 

\section{An Application to Colorectal Cancer}

We applied the naive (or conventional) and the five MR-based methods to assess the genetic interaction with the body mass index (BMI) in association with colorectal cancer (CRC) risk using the pooled data of studies from the Colon Cancer Family Registry, the Colorectal Transdisciplinary Study, the Genetics and Epidemiology of Colorectal Cancer Consortium, and the United Kingdom Biobank. The details of these studies have been previously published \citep{huyghe2019discovery, schmit2019novel,schumacher2015genome}. Briefly, the studies were either population-based case-control studies or nested case-control studies assembled from cohort studies via risk-set sampling. Cases and controls were matched on age, sex, race, and enrollment date or trial group, when applicable. Colorectal adenocarcinoma cases were confirmed by medical records, pathological reports, or death certificate information. All participants gave written informed consent and studies were approved by their respective Institutional Review Boards. 

Analyses were limited to individuals of European ancestry based on self-reported race and clustering of principal components with 1000 Genomes European population. Individuals were further excluded based on cryptic relatedness or duplicates (prioritizing cases and/or individuals genotyped on the better platform), genotyping or imputation errors, non-CRC outcomes, and age outliers. The final pooled sample size was 52,235 controls and 44,500 cases. 
Demographics and environmental exposures, here, BMI, were self-reported either at in-person interviews or via structured questionnaires and were harmonized across studies through a multi-step procedure \citep{hutter2012characterization}. All individuals were genotyped by microarrays and the quality control steps had been previously published \citep{huyghe2019discovery}. Briefly, genotyped SNPs were excluded based on call-rate ($< 95-98\%$), lack of Hardy-Weinberg equilibrium (p-value $< 10^{-4}$), discrepancies between reported and genotypic sex, discordant calls between duplicates. All autosomal SNPs were imputed to the Haplotype Reference Consortium r1.1 (2016) reference panel via the Michigan Imputation Server \citep{das2016next}. Imputed common SNPs were restricted based on a pooled MAF $\geq 5\%$ and imputation accuracy (R$^2>0.9$). After imputation and quality control analyses, a total of about 5.5 million common SNPs were included.  

To predict BMI, we fit a linear regression model including three PRS previously developed by \citep{yengo2018meta, kichaev2019leveraging} (1,368 SNPs), \cite{bull2020adiposity} (312 SNPs), and  \cite{pulit2019meta} (1,092 SNPs) using the 52,235 controls from the pooled dataset. The model also includes study, age, sex, and three principal components (PCs) for genetic ancestry. The R$^2$ explained by all predictors is 9.3\%. All three PRSs were significantly associated with BMI (p-values are $1.6\text{e-}4$, $1.4\text{e-}4$, and $<2\text{e-}16$, respectively) and the three PRSs alone explained 6.2\% of variation of BMI, which is considerably greater than typical biomarkers and lifestyle factors, see e.g., \cite{burgess2013use} and references therein. Based on this model, we constructed two scores for predicting BMI. The first overall score was the weighted sum of the three PRSs as well as other non-genetic predictors including age, sex, PCs, and study, with weights being the regression coefficient estimates in the linear model. The second score was based on only the three PRS. For the 2SPS and adjusted 2SPS estimators, we used the first overall score, and for the 2SPSa and adjusted 2SPSa we used the second PRS-only score. For the 2SRI, we calculated the residual by subtracting the first overall score from BMI. All models (naive and MR) were adjusted for study, age, sex, and three PCs.

\begin{figure}[H] 
    \centering
    \includegraphics[width=16cm,height=19cm]{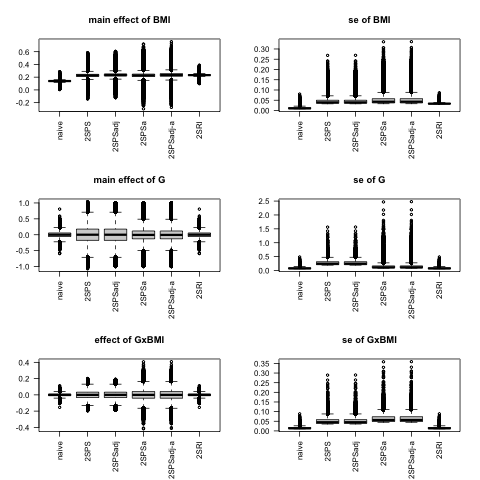}
    \caption{Boxplots of the estimates for the main effects of BMI and G, and the interaction GxBMI (left column), and the standard errors of corresponding estimates (right column) for the naive estimator and 5 MR estimators. 
    \label{fig:boxplot}}
\end{figure}

\begin{figure}[H]
    \centering
    \includegraphics[width=16cm,height=12cm]{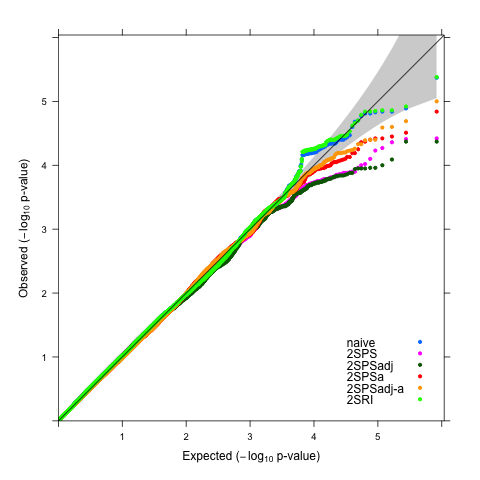}
    \caption{QQplots of $-\log_{10}\text{p-value}$ for GxBMI for all estimators versus the expected $-\log_{10}\text{p-value}$ under the uniform distribution. 
    } \label{fig:qqplot}
\end{figure}

\begin{figure}[H]
    \centering
    \includegraphics[width=16cm,height=17cm]{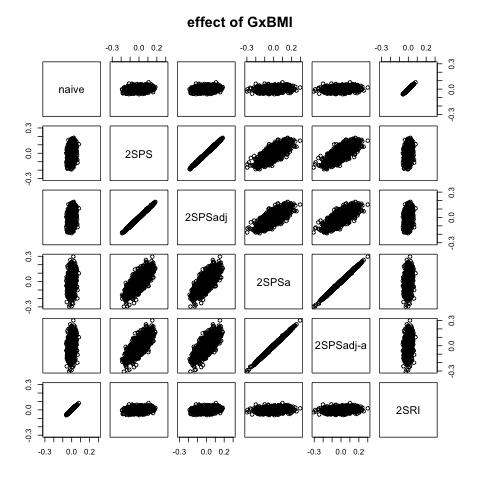}
    \caption{Pairwise plots of the estimates for GxBMI for all estimators. 
    } \label{fig:pairwise}
\end{figure}

\begin{figure}[H]
    \centering
    \includegraphics[width=16cm,height=17cm]{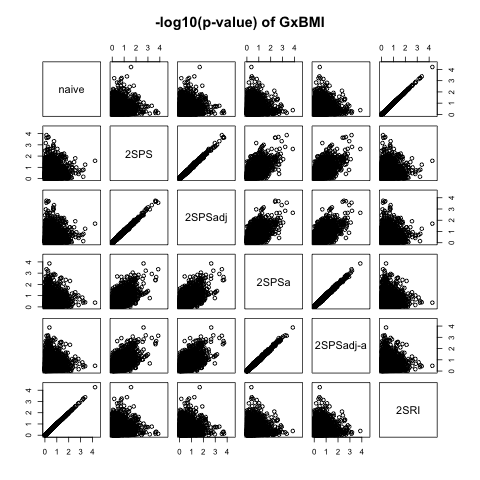}
    \caption{Pairwise plots of $-\log_{10}\text{p-value}$ for GxBMI for all estimators. 
    } \label{fig:pairwisepvalue}
\end{figure}

Figure~\ref{fig:boxplot} shows the boxplots of regression coefficient estimates and standard errors for the main effects of BMI (i.e., $X$) and $G$ and the interaction effect between BMI and $G$ for the naive and the five MR-based estimators. Here we focus on the results for Chromosome 1 (in each run $G$ is one of the 414,107 SNPs of Chromosome 1) to compare the relative performance of all estimators (the results for other chromosomes are similar). As it can be seen from Figure~\ref{fig:boxplot}, the estimates of interaction effect of GxBMI  centered around 0 for all estimators, suggesting that there was no overall GxBMI interaction. The QQplot that compared 
$-\log_{10}\text{p-value}$ with the expected under the uniform distribution also showed no evidence of inflation for GxBMI interaction for all estimators (Figure~\ref{fig:qqplot}). As majority of SNPs are not associated with CRC risk, this result is consistent with our simulation results under the main effect for $G$ being null where the type-I error for all estimators is generally maintained (Table~\ref{tbl:logisticMore}). The QQplot also shows that there was no evidence for GxBMI interaction under the overall null. The naive and 2SRI estimators had the smallest standard errors with mean $\sim 0.015$ for both estimators. The four variations of 2SPS estimators all have greater standard errors with 2SPSa and adjusted 2SPSa estimators having largest standard errors $\sim 0.063$ vs $\sim 0.050$, which may be due to the fact that the BMI predictor for the 2SPSa estimators based on only PRS was less predictive than that for the 2SPS estimators that included both the PRS and other non-genetic predictors. For the 2SPS and 2SPSa, adjusting for the residuals did not improve the efficiency. The pairwise plots also showed that the 2SPS and adjusted 2SPS were highly correlated in both the estimates and p-values, so were the 2SPSa and adjusted 2SPSa estimators (Figures~\ref{fig:pairwise} and \ref{fig:pairwisepvalue}). 2SPS and 2SPSa were also correlated, but not as strongly. The naive and 2SRI estimators were highly correlated in both the estimates and p-values. However, the 2SPS and 2SPSa estimators had little correlation with the naive and 2SRI estimators with the 2SPS and 2SPSa estimators having wider ranges, possibly because of their greater variances. 
The patterns for the main effects of $G$ and BMI were similar (Figure~\ref{fig:boxplot}, Figures~S1--S4 of SI). The notable difference is that when the main effect is non-zero, here, BMI, the p-values for various estimators were more correlated than when the main effect is null, here, $G$. We conducted a simulation study mimicking the real data and found the patterns observed for the real data were consistent to the simulated data (Figure~S5 of SI). 

It is worth noting that the main effect estimates of BMI based on all the methods had a non-zero median. While it is generally not meaningful to examine the main effects in the presence of an interaction effect, since there was no evidence for GxBMI interaction, we could evaluate the effect of BMI on CRC risk. All estimators suggested that higher BMI is associated with increased CRC risk. Interestingly, the log-odds ratio estimates were 0.228,  0.234,  0.228,  0.235, and 0.234 per 5 unit increase in BMI for the 2SPS, adjusted 2SPS, 2SPSa, adjusted 2SPSa, and 2SRI estimators, respectively, all of which were greater than the naive estimator, 0.139. This difference, especially the MR estimate greater than the naive estimator based on conventional observational studies, has been noted before \citep{suzuki2021body}. Possible explanations include inherent limitations of conventional observational studies such as reverse causation \citep{mandic2023association} and measurement error (i.e. residual confounding) \citep{davey2003mendelian}. For example, if a particular behavior is associated with increased risk but this behavior is also associated with under reporting BMI, failing to adjust for this behavior will result in attenuation of the effect of BMI. This is consistent with our simulation result when the effect of the unobserved confounder $U$ has opposite direction for the exposure and outcome (Figure~\ref{fig:logistic3Anegative}).  Other possible explanations may be that genetically determined BMI is likely to reflect lifelong exposure and there is empirical evidence showing that the magnitude of the association for lifelong exposure for the risk of disease is larger than that for the short-term exposure beginning later in life \citep{ference2012effect}. Another possibility is that the MR estimates might have been inflated, see for example, Figure \ref{fig:logistic3Anegative}; however, such possibility seems to be less likely, in part because the inflation, if exists, limits to the 2SPS estimators not the 2SRI estimator and in our real data analysis, 2SPS and 2SRI estimates were consistent and they were all greater than the naive estimate.  

\section{Discussion}
This research focuses on practical MR models designed for both testing and estimating the effect of GxE on $Y$. Extensions of the 2SPS and 2SRI methods tailored for linear and logistic regression models of the outcome variable $Y$ have been introduced and extensively evaluated through a simulation study. In the case of the linear model, all the MR-based approaches, as well as the naive approach, provide a valid test for GxE. However, the naive estimators often exhibit bias. Among the MR-based methods, 2SRI is recommended for mitigating the bias associated with the naive approach, as it is shown to be the most efficient approach. Regarding the logistic regression model,  the MR-based methods generally maintains type-I error with limited inflation, when the effect of unobserved confounder $U$ on $Y$ is weak or when there is no measured confounder $Z$. All estimators were biased, though the 2SRI estimator has the smallest bias, especially for the main effect of the exposure, which can be substantial under the naive method. This was demonstrated by our application to colorectal study, where, in fact, the effect of observed BMI was substantially attenuated compared to the MR estimators. 

Our research can be expanded in several directions. First, we have considered perhaps the simplest model for the unmeasured confounder. In practice, it is likely that the unmeasured confounder  interacts with the exposure or with $G$. Then, some of the current observations may no longer hold. Second, in the real data analysis, three PRSs were used, each involving hundreds to over a thousand genetic variants, as genetic instruments for BMI in order to increase the power of the MR analysis \citep{burgess2013use}. However, it is likely that some of the genetic variants in the PRSs may not be valid IVs, in the sense that they may have direct effects on the outcome, beside mediating through the BMI. Including invalid IVs may bias the results. Many methods have been developed to relax this assumption by incorporating a horizontal pleiotropy effect \citep{burgess2017interpreting}, allowing for some direct effects under the InSide assumption \citep{bowden1984instrumental}, or identifying invalid IVs \citep{verbanck2018detection}, but not in the context of gene-environment interaction.   Third, the popular logistic regression model for binary outcome clearly has limitation when it comes to the MR analysis due to its non-collapsibility. Due to lack of a close form or approximation of a close form for the commonly used MR estimators, especially the 2SPS-type estimators, it is not easy to understand the magnitude and direction of the biases based on entirely empirical results, the complexity of which have been shown in our extensive simulation study. Future research towards better understanding of the MR methods for binary outcome is needed. 

As  put by \cite{vanderweele2009distinction}, it is important to distinguish between interaction and effect modification. In particular, 
``interaction is defined in terms of the effects of two interventions whereas effect modification is defined in terms of the effect of one intervention varying across strata of a second variable''.
Under the above effect-modification definition, there is asymmetry between $G$ and $X$.  The role of $G$ concerns whether the effect of primary interest varies across strata defined by $G$. Effect modification can have important implications for public health as the impact of  the exposure $X$  may differ across sub-populations defined by $G$. This may indicate the necessity for different interventions among these sub-populations. Based on causal DAGs and Rule 2 of Pearl's do-calculus \citep{pearl1995causal}, \cite{vanderweele2009distinction} provided the required condition for interaction and effect modification to coincide. Hence, some of the results presented in this work are also applicable in effect modification formulation, although, methods directly targeting estimands of effect modification could be more efficient, and are beyond the scope of this work.

\section{Acknowledgements}
The work is supported in part by grants from the National Institutes of Health (R01CA189532, R01HL145806, U01 CA164930, R01 CA20140), the Israel Science Foundation (ISF) grant number 767/21 and by a grant from the Tel-Aviv University Center for AI and Data Science (TAD). The authors thank the Genetics and Epidemiology of Colorectal Cancer Consortium (GECCO) and the participating studies, investigators, staff, and study participants for their dedication and contributions.  A full list of these studies and funding is provided in the Supporting Information.

\bibliography{literature}


\clearpage

\renewcommand{\baselinestretch}{1} 

\begin{table}[H]
\scriptsize
\centering
\caption{Summary of simulation settings}\label{tbl:sim-desc}
\begin{tabular}{lcccc}
  \hline
Setting & Model of $Y$ & Value of $\beta_3$ & $G$ and $G_{IV}$ \\ 
     \hline
     I.A & linear & 0.5  & independent \\
     I.B & linear & 0.0  & independent \\
     II.A & linear & 0.5  & dependent \\
     II.B & linear & 0.0  & dependent \\
     III.A & logistic & 0.5  & independent \\
     III.B & logistic & 0.0  & independent \\
     IV.A & logistic & 0.5  & dependent \\
     IV.B & logistic & 0.0  & dependent \\     
     \hline
\end{tabular}
\end{table}

\begin{table}[H]
\scriptsize
\centering
\caption{Simulation results, linear regression, Setting I.B - Type I error of $H_0: \beta_3=0$ vs a two-sided alternative, based on 500 simulated datasets.
}\label{tbl:GXIblinear}
\begin{tabular}{rrrrrrrrrrr}
  \hline
 & \multicolumn{2}{c}{$\gamma_U=0$} & \multicolumn{2}{c}{$\gamma_U=0.5$} &
 \multicolumn{2}{c}{$\gamma_U=1$} & \multicolumn{2}{c}{$\gamma_U=2$}  & \multicolumn{2}{c}{$\gamma_U=4$}\\ 
 & Mean  &SD & Mean & SD & Mean & SD & Mean & SD & Mean & SD \\ 
  \hline
   \multicolumn{11}{c}{$\beta_U=0$} \\
Naive  & 0.04 & 0.20 & 0.04 & 0.20 & 0.05 & 0.21 & 0.06 & 0.23 & 0.05 & 0.21 \\ 
2SPS  & 0.05 & 0.21 & 0.06 & 0.23 & 0.06 & 0.24 & 0.07 & 0.26 & 0.07 & 0.26 \\ 
  2SPSadj  & 0.04 & 0.19 & 0.04 & 0.19 & 0.04 & 0.19 & 0.04 & 0.19 & 0.03 & 0.18 \\ 
  2SPSa & 0.05 & 0.21 & 0.05 & 0.21 & 0.05 & 0.21 & 0.04 & 0.20 & 0.04 & 0.20 \\ 
  2SPSadj-a & 0.05 & 0.22 & 0.05 & 0.21 & 0.05 & 0.21 & 0.05 & 0.21 & 0.05 & 0.21 \\ 
  2SRI  & 0.04 & 0.20 & 0.04 & 0.20 & 0.05 & 0.21 & 0.06 & 0.23 & 0.05 & 0.21 \\ 
     \multicolumn{11}{c}{$\beta_U=1.5$} \\
  Naive  & 0.05 & 0.23 & 0.06 & 0.24 & 0.04 & 0.21 & 0.03 & 0.18 & 0.05 & 0.21 \\ 
  2SPS  & 0.07 & 0.25 & 0.07 & 0.25 & 0.07 & 0.26 & 0.08 & 0.27 & 0.07 & 0.26 \\ 
  2SPSadj  & 0.07 & 0.26 & 0.06 & 0.24 & 0.05 & 0.21 & 0.04 & 0.20 & 0.04 & 0.19 \\ 
  2SPSa & 0.04 & 0.20 & 0.04 & 0.21 & 0.04 & 0.20 & 0.04 & 0.20 & 0.05 & 0.21 \\ 
  2SPSadj-a & 0.05 & 0.21 & 0.06 & 0.23 & 0.05 & 0.23 & 0.05 & 0.21 & 0.04 & 0.20 \\ 
  2SRI  & 0.05 & 0.23 & 0.06 & 0.23 & 0.04 & 0.21 & 0.04 & 0.19 & 0.05 & 0.21 \\ 
     \multicolumn{11}{c}{$\beta_U=3$} \\
      Naive  & 0.04 & 0.20 & 0.05 & 0.22 & 0.06 & 0.23 & 0.05 & 0.22 & 0.05 & 0.21 \\ 
  2SPS  & 0.07 & 0.26 & 0.07 & 0.26 & 0.06 & 0.24 & 0.06 & 0.24 & 0.07 & 0.25 \\ 
  2SPSadj  & 0.07 & 0.25 & 0.06 & 0.23 & 0.06 & 0.24 & 0.05 & 0.22 & 0.04 & 0.19 \\ 
  2SPSa & 0.05 & 0.22 & 0.05 & 0.21 & 0.05 & 0.21 & 0.05 & 0.21 & 0.05 & 0.21 \\ 
  2SPSadj-a & 0.05 & 0.22 & 0.06 & 0.24 & 0.06 & 0.24 & 0.05 & 0.21 & 0.04 & 0.21 \\ 
  2SRI  & 0.04 & 0.20 & 0.05 & 0.23 & 0.06 & 0.23 & 0.05 & 0.21 & 0.05 & 0.21 \\ 
     \hline
\end{tabular}
\end{table}

\begin{table}[H]
\scriptsize
\centering
\caption{Simulation results, linear regression, Setting II.B - Type I error of $H_0: \beta_3=0$ vs a two-sided alternative, based on 500 simulated datasets. 
}\label{tbl:GXIIblinear}
\begin{tabular}{rrrrrrrrrrr}
  \hline
 & \multicolumn{2}{c}{$\gamma_U=0$} & \multicolumn{2}{c}{$\gamma_U=0.5$} &
 \multicolumn{2}{c}{$\gamma_U=1$} & \multicolumn{2}{c}{$\gamma_U=2$}  & \multicolumn{2}{c}{$\gamma_U=4$}\\ 
 & Mean  &SD & Mean & SD & Mean & SD & Mean & SD & Mean & SD \\ 
  \hline
   \multicolumn{11}{c}{$\beta_U=0$} \\
Naive  & 0.05 & 0.21 & 0.05 & 0.21 & 0.06 & 0.23 & 0.06 & 0.23 & 0.06 & 0.23 \\  
  2SPS  & 0.05 & 0.22 & 0.06 & 0.24 & 0.06 & 0.23 & 0.05 & 0.21 & 0.05 & 0.21 \\  
  2SPSadj  & 0.07 & 0.26 & 0.07 & 0.26 & 0.07 & 0.26 & 0.07 & 0.26 & 0.07 & 0.26 \\ 
  2SPSa & 0.05 & 0.21 & 0.04 & 0.21 & 0.04 & 0.20 & 0.05 & 0.22 & 0.06 & 0.23 \\ 
  2SPSadj-a & 0.05 & 0.22 & 0.05 & 0.23 & 0.05 & 0.23 & 0.05 & 0.22 & 0.05 & 0.22 \\ 
  2SRI  & 0.05 & 0.21 & 0.05 & 0.21 & 0.06 & 0.23 & 0.06 & 0.23 & 0.06 & 0.23 \\ 
\multicolumn{11}{c}{$\beta_U=1.5$} \\
 Naive  & 0.03 & 0.18 & 0.04 & 0.19 & 0.05 & 0.22 & 0.05 & 0.23 & 0.06 & 0.23 \\ 
  2SPS  & 0.06 & 0.23 & 0.06 & 0.24 & 0.06 & 0.24 & 0.05 & 0.23 & 0.06 & 0.23 \\ 
  2SPSadj & 0.04 & 0.20 & 0.05 & 0.21 & 0.05 & 0.23 & 0.06 & 0.23 & 0.06 & 0.24 \\ 
  2SPSa & 0.05 & 0.22 & 0.05 & 0.22 & 0.05 & 0.22 & 0.04 & 0.21 & 0.04 & 0.21 \\ 
 2SPSadj-a & 0.04 & 0.20 & 0.04 & 0.19 & 0.04 & 0.20 & 0.06 & 0.23 & 0.06 & 0.23 \\ 
  2SRI  & 0.03 & 0.18 & 0.04 & 0.19 & 0.05 & 0.21 & 0.06 & 0.23 & 0.06 & 0.23 \\ 
   \multicolumn{11}{c}{$\beta_U=3$} \\
 Naive  & 0.04 & 0.19 & 0.04 & 0.19 & 0.05 & 0.22 & 0.04 & 0.21 & 0.06 & 0.23 \\ 
  2SPS  & 0.05 & 0.22 & 0.05 & 0.22 & 0.05 & 0.23 & 0.05 & 0.22 & 0.05 & 0.23 \\ 
  2SPSadj  & 0.04 & 0.20 & 0.04 & 0.19 & 0.05 & 0.21 & 0.05 & 0.22 & 0.06 & 0.23 \\ 
  2SPS-a & 0.05 & 0.22 & 0.05 & 0.21 & 0.04 & 0.20 & 0.04 & 0.21 & 0.04 & 0.20 \\ 
  2SPSadj-a & 0.04 & 0.21 & 0.04 & 0.19 & 0.03 & 0.18 & 0.05 & 0.21 & 0.05 & 0.23 \\ 
  2SRI  & 0.04 & 0.19 & 0.04 & 0.19 & 0.05 & 0.21 & 0.05 & 0.21 & 0.05 & 0.21 \\    
     \hline
\end{tabular}
\end{table}


\begin{table}[H]
\scriptsize
\centering
\caption{Simulation results, logistic regression, Setting III.B - Type I error of $H_0: \beta_3=0$ vs a two-sided alternative, based on 500 simulated datasets. 
}\label{tbl:logistic3B}
\begin{tabular}{rrrrrrrrrrr}
  \hline
 & \multicolumn{2}{c}{$\gamma_U=0$} & \multicolumn{2}{c}{$\gamma_U=0.5$} &
 \multicolumn{2}{c}{$\gamma_U=1$} & \multicolumn{2}{c}{$\gamma_U=2$}  & \multicolumn{2}{c}{$\gamma_U=4$}\\ 
 & Mean  &SD & Mean & SD & Mean & SD & Mean & SD & Mean & SD \\ 
  \hline
   \multicolumn{11}{c}{$\beta_U=0$} \\
Naive & 0.06 & 0.23 & 0.06 & 0.24 & 0.04 & 0.21 & 0.06 & 0.23 & 0.05 & 0.22 \\ 
2SPS & 0.06 & 0.24 & 0.06 & 0.24 & 0.07 & 0.26 & 0.08 & 0.27 & 0.07 & 0.26 \\ 
2SPSadj & 0.06 & 0.23 & 0.06 & 0.23 & 0.05 & 0.21 & 0.06 & 0.24 & 0.07 & 0.25 \\ 
2SPSa & 0.07 & 0.26 & 0.07 & 0.26 & 0.06 & 0.23 & 0.07 & 0.26 & 0.08 & 0.27 \\ 
2SPSadj-a & 0.06 & 0.24 & 0.06 & 0.24 & 0.06 & 0.23 & 0.05 & 0.22 & 0.05 & 0.23 \\ 
2SRI & 0.05 & 0.23 & 0.06 & 0.24 & 0.04 & 0.21 & 0.06 & 0.23 & 0.05 & 0.22 \\  
     \multicolumn{11}{c}{$\beta_U=1.5$} \\
Naive & 0.09 & 0.28 & 0.07 & 0.26 & 0.07 & 0.26 & 0.07 & 0.25 & 0.04 & 0.20 \\ 
  2SPS  & 0.09 & 0.29 & 0.10 & 0.30 & 0.07 & 0.26 & 0.09 & 0.28 & 0.07 & 0.26 \\ 
  2SPSadj & 0.06 & 0.24 & 0.04 & 0.20 & 0.03 & 0.18 & 0.05 & 0.22 & 0.04 & 0.19 \\ 
  2SPS-a & 0.05 & 0.23 & 0.05 & 0.23 & 0.06 & 0.23 & 0.06 & 0.23 & 0.04 & 0.21 \\  
  2SPSadj-a & 0.06 & 0.23 & 0.04 & 0.19 & 0.05 & 0.21 & 0.04 & 0.21 & 0.05 & 0.21 \\ 
  2SRI & 0.08 & 0.28 & 0.07 & 0.26 & 0.06 & 0.24 & 0.07 & 0.25 & 0.04 & 0.19 \\
     \multicolumn{11}{c}{$\beta_U=3$} \\
      Naive & 0.08 & 0.27 & 0.08 & 0.27 & 0.05 & 0.22 & 0.04 & 0.20 & 0.07 & 0.25 \\
  2SPS & 0.04 & 0.20 & 0.06 & 0.23 & 0.07 & 0.26 & 0.05 & 0.22 & 0.05 & 0.23 \\ 
  2SPSadj & 0.04 & 0.19 & 0.04 & 0.21 & 0.06 & 0.23 & 0.05 & 0.21 & 0.06 & 0.24 \\ 
  2SPS-a & 0.04 & 0.20 & 0.06 & 0.23 & 0.05 & 0.22 & 0.03 & 0.18 & 0.05 & 0.21 \\ 
  2SPSadj-a & 0.04 & 0.19 & 0.07 & 0.26 & 0.04 & 0.20 & 0.03 & 0.18 & 0.04 & 0.21 \\ 
  2SRI  & 0.08 & 0.27 & 0.07 & 0.26 & 0.05 & 0.22 & 0.05 & 0.21 & 0.05 & 0.21 \\ 
     \hline
\end{tabular}
\end{table}

\begin{table}[H]
\scriptsize
\centering
\caption{Simulation results, logistic regression, Setting IV.B - Type I error of $H_0: \beta_3=0$ vs a two-sided alternative, based on 500 simulated datasets. 
}\label{tbl:logistic4B}
\begin{tabular}{rrrrrrrrrrr}
  \hline
 & \multicolumn{2}{c}{$\gamma_U=0$} & \multicolumn{2}{c}{$\gamma_U=0.5$} &
 \multicolumn{2}{c}{$\gamma_U=1$} & \multicolumn{2}{c}{$\gamma_U=2$}  & \multicolumn{2}{c}{$\gamma_U=4$}\\ 
 & Mean  &SD & Mean & SD & Mean & SD & Mean & SD & Mean & SD \\ 
  \hline
   \multicolumn{11}{c}{$\beta_U=0$} \\
 Naive  & 0.06 & 0.24 & 0.07 & 0.25 & 0.06 & 0.24 & 0.05 & 0.21 & 0.04 & 0.20 \\ 
  2SPS & 0.07 & 0.25 & 0.07 & 0.25 & 0.05 & 0.21 & 0.07 & 0.25 & 0.05 & 0.23 \\ 
  2SPSadj  & 0.05 & 0.23 & 0.04 & 0.21 & 0.04 & 0.21 & 0.04 & 0.19 & 0.04 & 0.19 \\ 
  2SPSa & 0.06 & 0.24 & 0.05 & 0.22 & 0.05 & 0.22 & 0.04 & 0.20 & 0.05 & 0.23 \\ 
  2SPSadj-a & 0.05 & 0.21 & 0.05 & 0.22 & 0.04 & 0.21 & 0.04 & 0.19 & 0.05 & 0.23 \\ 
  2SRI  & 0.06 & 0.23 & 0.07 & 0.26 & 0.06 & 0.24 & 0.05 & 0.21 & 0.04 & 0.20 \\
  \multicolumn{11}{c}{$\beta_U=1.5$} \\
  Naive  & 0.15 & 0.36 & 0.12 & 0.32 & 0.06 & 0.24 & 0.06 & 0.23 & 0.04 & 0.20 \\ 
  2SPS  & 0.13 & 0.34 & 0.14 & 0.35 & 0.13 & 0.33 & 0.11 & 0.31 & 0.09 & 0.28 \\ 
  2SPSadj & 0.09 & 0.29 & 0.08 & 0.27 & 0.05 & 0.21 & 0.05 & 0.22 & 0.05 & 0.22 \\ 
  2SPSa & 0.08 & 0.27 & 0.06 & 0.24 & 0.06 & 0.24 & 0.05 & 0.21 & 0.05 & 0.22 \\ 
  2SPSadj-a & 0.06 & 0.23 & 0.05 & 0.21 & 0.06 & 0.24 & 0.05 & 0.23 & 0.04 & 0.19 \\ 
  2SRI  & 0.16 & 0.36 & 0.14 & 0.35 & 0.07 & 0.25 & 0.05 & 0.21 & 0.04 & 0.21 \\ 
  \multicolumn{11}{c}{$\beta_U=3$} \\
   Naive  & 0.11 & 0.31 & 0.10 & 0.30 & 0.06 & 0.24 & 0.05 & 0.23 & 0.05 & 0.23 \\ 
  2SPS  & 0.09 & 0.28 & 0.09 & 0.28 & 0.06 & 0.23 & 0.06 & 0.23 & 0.05 & 0.22 \\ 
  2SPSadj & 0.07 & 0.26 & 0.06 & 0.24 & 0.05 & 0.23 & 0.04 & 0.19 & 0.04 & 0.20 \\ 
  2SPSa & 0.04 & 0.21 & 0.07 & 0.25 & 0.05 & 0.21 & 0.05 & 0.23 & 0.05 & 0.23 \\ 
  2SPSadj-a & 0.04 & 0.21 & 0.05 & 0.23 & 0.04 & 0.20 & 0.05 & 0.22 & 0.05 & 0.22 \\
  2SRI  & 0.11 & 0.31 & 0.11 & 0.32 & 0.08 & 0.27 & 0.06 & 0.24 & 0.05 & 0.21 \\ 
     \hline
\end{tabular}
\end{table}

\begin{table}[H]
\scriptsize
\centering
\caption{Simulation results, logistic regression, $n=30,000$ - Type I error of $H_0: \beta_3=0$ vs a two-sided alternative, based on 1000 simulated datasets. 
}\label{tbl:logisticMore}
\begin{tabular}{rrrrrrrrrrr}
  \hline
 & \multicolumn{2}{c}{$\gamma_U=0$} & \multicolumn{2}{c}{$\gamma_U=0.5$} &
 \multicolumn{2}{c}{$\gamma_U=1$} & \multicolumn{2}{c}{$\gamma_U=2$}  & \multicolumn{2}{c}{$\gamma_U=4$}\\ 
 & Mean  &SD & Mean & SD & Mean & SD & Mean & SD & Mean & SD \\ 
  \hline
  \multicolumn{11}{c}{$\beta_U=0.5$, $\beta_Z=\gamma_Z=0$, $\beta_2=0.5$  } \\
  Naive  & 0.06 & 0.23 & 0.06 & 0.24 & 0.05 & 0.22 & 0.06 & 0.23 & 0.05 & 0.22 \\ 
  2SPS & 0.07 & 0.26 & 0.08 & 0.28 & 0.07 & 0.25 & 0.05 & 0.22 & 0.05 & 0.21 \\ 
  2SPSadj  & 0.06 & 0.23 & 0.06 & 0.23 & 0.05 & 0.21 & 0.05 & 0.22 & 0.05 & 0.21 \\ 
  2SPSa & 0.07 & 0.26 & 0.09 & 0.28 & 0.07 & 0.25 & 0.05 & 0.21 & 0.05 & 0.22 \\ 
  2SPSadj-a & 0.06 & 0.24 & 0.06 & 0.23 & 0.05 & 0.21 & 0.05 & 0.22 & 0.05 & 0.21 \\ 
  2SRI & 0.06 & 0.23 & 0.06 & 0.23 & 0.05 & 0.21 & 0.06 & 0.23 & 0.05 & 0.22 \\ 
   \multicolumn{11}{c}{$\beta_U=3$, $\beta_Z=\gamma_Z=0$, $\beta_2=0.5$  } \\
Naive  & 0.13 & 0.33 & 0.14 & 0.35 & 0.11 & 0.32 & 0.07 & 0.25 & 0.05 & 0.21 \\ 
  2SPS  & 0.05 & 0.22 & 0.06 & 0.23 & 0.07 & 0.26 & 0.04 & 0.20 & 0.04 & 0.20 \\ 
  2SPSadj  & 0.05 & 0.21 & 0.05 & 0.21 & 0.05 & 0.21 & 0.05 & 0.21 & 0.04 & 0.19 \\ 
  2SPSa & 0.05 & 0.23 & 0.06 & 0.24 & 0.07 & 0.26 & 0.04 & 0.20 & 0.04 & 0.20 \\ 
  2SPSadj-a  & 0.05 & 0.21 & 0.05 & 0.22 & 0.05 & 0.22 & 0.05 & 0.21 & 0.04 & 0.20 \\ 
  2SRI  & 0.13 & 0.34 & 0.13 & 0.33 & 0.10 & 0.30 & 0.06 & 0.24 & 0.05 & 0.22 \\ 
    \multicolumn{11}{c}{$\beta_U=0.5$, $\beta_Z=\gamma_Z=0.5$, $\beta_2=0.5$  } \\
  Naive  & 0.06 & 0.24 & 0.06 & 0.24 & 0.06 & 0.23 & 0.05 & 0.22 & 0.06 & 0.24 \\ 
  2SPS  & 0.09 & 0.28 & 0.11 & 0.31 & 0.12 & 0.32 & 0.11 & 0.31 & 0.09 & 0.28 \\ 
  2SPSadj  & 0.07 & 0.25 & 0.05 & 0.22 & 0.04 & 0.20 & 0.05 & 0.21 & 0.05 & 0.21 \\ 
  2SPSa & 0.06 & 0.24 & 0.07 & 0.25 & 0.07 & 0.25 & 0.06 & 0.24 & 0.05 & 0.21 \\ 
  2SPSadj-a & 0.05 & 0.22 & 0.06 & 0.23 & 0.05 & 0.21 & 0.06 & 0.23 & 0.04 & 0.21 \\ 
  2SRI  & 0.06 & 0.24 & 0.06 & 0.24 & 0.06 & 0.23 & 0.05 & 0.23 & 0.06 & 0.24 \\ 
   \multicolumn{11}{c}{$\beta_U=3$, $\beta_Z=\gamma_Z=0.5$, $\beta_2=0.5$ } \\
  Naive  & 0.15 & 0.35 & 0.14 & 0.35 & 0.10 & 0.30 & 0.06 & 0.24 & 0.04 & 0.20 \\ 
  2SPS  & 0.10 & 0.31 & 0.09 & 0.28 & 0.08 & 0.27 & 0.07 & 0.26 & 0.05 & 0.22 \\ 
  2SPSadj  & 0.09 & 0.29 & 0.05 & 0.22 & 0.05 & 0.22 & 0.04 & 0.19 & 0.07 & 0.25 \\ 
  2SPSa & 0.07 & 0.25 & 0.07 & 0.26 & 0.06 & 0.24 & 0.04 & 0.21 & 0.04 & 0.21 \\ 
  2SPSadj-a & 0.07 & 0.25 & 0.05 & 0.22 & 0.07 & 0.25 & 0.05 & 0.22 & 0.05 & 0.23 \\ 
  2SRI  & 0.15 & 0.35 & 0.13 & 0.34 & 0.10 & 0.30 & 0.06 & 0.23 & 0.04 & 0.20 \\ 
   \multicolumn{11}{c}{$\beta_U=3$, $\beta_Z=\gamma_Z=0.5$, $\beta_2=0$ } \\
  Naive & 0.04 & 0.20 & 0.05 & 0.21 & 0.05 & 0.21 & 0.05 & 0.23 & 0.05 & 0.22 \\ 
  2SPS  & 0.04 & 0.20 & 0.05 & 0.22 & 0.04 & 0.20 & 0.05 & 0.22 & 0.06 & 0.24 \\ 
  2SPSadj  & 0.05 & 0.21 & 0.06 & 0.23 & 0.05 & 0.23 & 0.05 & 0.21 & 0.05 & 0.22 \\ 
  2SPSa & 0.04 & 0.20 & 0.04 & 0.21 & 0.04 & 0.19 & 0.04 & 0.20 & 0.06 & 0.23 \\ 
  2SPSadj-a & 0.04 & 0.20 & 0.04 & 0.20 & 0.05 & 0.22 & 0.06 & 0.24 & 0.06 & 0.23 \\ 
  2SRI  & 0.04 & 0.20 & 0.05 & 0.23 & 0.05 & 0.22 & 0.06 & 0.23 & 0.05 & 0.23 \\ 
       \hline
\end{tabular}
\end{table}

\newpage

\newpage

\clearpage

\clearpage

\clearpage

\clearpage

\renewcommand\thesection{S\arabic{section}}
\renewcommand{\thetable}{S\arabic{table}}
\renewcommand{\thefigure}{S\arabic{figure}}
\renewcommand{\theequation}{S.\arabic{equation}}

\setcounter{table}{0}
\setcounter{figure}{0}

\section{Supporting Information}

\begin{table}[ht]
	\scriptsize
	\centering
	\caption{Simulation results, linear regression, Setting I.A} 
\begin{tabular}{rrrrrrrrrrr}
		\hline
		& \multicolumn{2}{c}{$\gamma_U=0$} & \multicolumn{2}{c}{$\gamma_U=0.5$} &
		\multicolumn{2}{c}{$\gamma_U=1$} & \multicolumn{2}{c}{$\gamma_U=2$}  & \multicolumn{2}{c}{$\gamma_U=4$}\\ 
		& Mean  &SD & Mean & SD & Mean & SD & Mean & SD & Mean & SD \\ 
		\hline
		$R^2_{X|G_{IV},Z,U}$ & 0.33 & 0.01 & 0.43 & 0.01 & 0.60 & 0.01 & 0.82 & 0.00 & 0.94 & 0.00 \\   
		$R^2_{X|G_{IV},Z}$  & 0.33 & 0.01 & 0.29 & 0.01 & 0.20 & 0.01 & 0.09 & 0.01 & 0.03 & 0.00 \\ 
		\multicolumn{11}{c}{$\beta_U=0$  } \\
		$R^2_{Y|X,G,Z,U}$  & 0.73 & 0.00 & 0.73 & 0.00 & 0.72 & 0.00 & 0.71 & 0.00 & 0.70 & 0.00 \\  
		$R^2_{Y|X,G,Z}$  & 0.73 & 0.00 & 0.73 & 0.00 & 0.72 & 0.00 & 0.71 & 0.00 & 0.70 & 0.00 \\  
		Naive $\beta_1$ & 1.00 & 0.01 & 1.00 & 0.01 & 1.00 & 0.01 & 1.00 & 0.01 & 1.00 & 0.01 \\
\hline
	\end{tabular}
\end{table}

\begin{table}[ht]
	\scriptsize
	\centering
	\caption{Simulation results, linear regression, Setting II.A}
	\begin{tabular}{rrrrrrrrrrr}
		\hline
		& \multicolumn{2}{c}{$\gamma_U=0$} & \multicolumn{2}{c}{$\gamma_U=0.5$} &
		\multicolumn{2}{c}{$\gamma_U=1$} & \multicolumn{2}{c}{$\gamma_U=2$} &
		\multicolumn{2}{c}{$\gamma_U=4$} \\ 
		& Mean  &SD & Mean & SD & Mean & SD & Mean & SD & Mean & SD  \\ 
		\hline
		$R^2_{X|G_{IV},Z,U}$ & 0.33 & 0.01 & 0.43 & 0.01 & 0.60 & 0.01 & 0.82 & 0.00 & 0.94 & 0.00 \\ 
		$R^2_{X|G_{IV},Z}$ & 0.33 & 0.01 & 0.29 & 0.01 & 0.20 & 0.01 & 0.09 & 0.01 & 0.03 & 0.00 \\ 
		\multicolumn{11}{c}{$\beta_U=0$} \\
		$R^2_{Y|X,G,Z,U}$ & 0.77 & 0.00 & 0.77 & 0.00 & 0.76 & 0.00 & 0.75 & 0.00 & 0.74 & 0.00 \\ 
		$R^2_{Y|X,G,Z}$ & 0.77 & 0.00 & 0.77 & 0.00 & 0.76 & 0.00 & 0.75 & 0.00 & 0.74 & 0.00 \\ 
		Naive $\beta_1$ & 1.00 & 0.02 & 1.00 & 0.02 & 1.00 & 0.02 & 1.00 & 0.02 & 1.00 & 0.02 \\
\hline
	\end{tabular}
\end{table}

\begin{table}[ht]
	\scriptsize
	\centering
	\caption{Simulation results, linear regression, Setting I.B}
	\begin{tabular}{rrrrrrrrrrr}
		\hline
		& \multicolumn{2}{c}{$\gamma_U=0$} & \multicolumn{2}{c}{$\gamma_U=0.5$} &
		\multicolumn{2}{c}{$\gamma_U=1$} & \multicolumn{2}{c}{$\gamma_U=2$}  & \multicolumn{2}{c}{$\gamma_U=4$}\\ 
		& Mean  &SD & Mean & SD & Mean & SD & Mean & SD & Mean & SD \\ 
		\hline
		$R^2_{X|G_{IV},Z,U}$ & 0.33 & 0.01 & 0.43 & 0.01 & 0.60 & 0.01 & 0.82 & 0.00 & 0.94 & 0.00 \\ 
		$R^2_{X|G_{IV},Z}$ & 0.33 & 0.01 & 0.29 & 0.01 & 0.20 & 0.01 & 0.09 & 0.01 & 0.03 & 0.00 \\  
		\multicolumn{11}{c}{$\beta_U=0$} \\
		$R^2_{Y|X,G,Z,U}$ & 0.64 & 0.01 & 0.63 & 0.01 & 0.63 & 0.01 & 0.61 & 0.01 & 0.60 & 0.01 \\ 
		$R^2_{Y|X,G,Z}$ & 0.64 & 0.01 & 0.63 & 0.01 & 0.63 & 0.01 & 0.61 & 0.01 & 0.60 & 0.01 \\ 
		Naive $\beta_1$ & 1.00 & 0.01 & 1.00 & 0.01 & 1.00 & 0.01 & 1.00 & 0.01 & 1.00 & 0.01 \\
\hline
	\end{tabular}
\end{table}

\begin{table}[ht]
	\scriptsize
	\centering
	\caption{Simulation results, linear regression, Setting II.B}
	\begin{tabular}{rrrrrrrrrrr}
		\hline
		& \multicolumn{2}{c}{$\gamma_U=0$} & \multicolumn{2}{c}{$\gamma_U=0.5$} &
		\multicolumn{2}{c}{$\gamma_U=1$} & \multicolumn{2}{c}{$\gamma_U=2$}  & \multicolumn{2}{c}{$\gamma_U=4$}\\ 
		& Mean  &SD & Mean & SD & Mean & SD & Mean & SD & Mean & SD \\ 
		\hline
		$R^2_{X|G_{IV},Z,U}$ & 0.33 & 0.01 & 0.43 & 0.01 & 0.60 & 0.01 & 0.82 & 0.00 & 0.94 & 0.00 \\ 
		$R^2_{X|G_{IV},Z}$ & 0.33 & 0.01 & 0.29 & 0.01 & 0.20 & 0.01 & 0.09 & 0.01 & 0.03 & 0.00 \\ 
		\multicolumn{11}{c}{$\beta_U=0$} \\
		$R^2_{Y|X,G,Z,U}$ & 0.65 & 0.00 & 0.65 & 0.00 & 0.64 & 0.00 & 0.62 & 0.01 & 0.61 & 0.01 \\ 
		$R^2_{Y|X,G,Z}$ & 0.65 & 0.00 & 0.65 & 0.00 & 0.64 & 0.00 & 0.62 & 0.01 & 0.60 & 0.01 \\ 
		Naive $\beta_1$ & 1.00 & 0.02 & 1.00 & 0.02 & 1.00 & 0.02 & 1.00 & 0.02 & 1.00 & 0.02 \\
\hline
	\end{tabular}
\end{table}


\begin{table}[ht]
	\scriptsize
	\centering
	\caption{Simulation results, logistic regression, Setting III.A}
	\begin{tabular}{rrrrrrrrrrr}
		\hline
		& \multicolumn{2}{c}{$\gamma_U=0$} & \multicolumn{2}{c}{$\gamma_U=0.5$} &
		\multicolumn{2}{c}{$\gamma_U=1$} & \multicolumn{2}{c}{$\gamma_U=2$}  & \multicolumn{2}{c}{$\gamma_U=4$}\\ 
		& Mean  &SD & Mean & SD & Mean & SD & Mean & SD & Mean & SD \\ 
		\hline
		$R^2_{X|G_{IV},Z,U}$ & 0.33 & 0.01 & 0.43 & 0.01 & 0.60 & 0.01 & 0.82 & 0.00 & 0.94 & 0.00 \\ 
		$R^2_{X|G_{IV},Z}$ & 0.33 & 0.01 & 0.29 & 0.01 & 0.20 & 0.01 & 0.09 & 0.01 & 0.03 & 0.00 \\ 
		\multicolumn{11}{c}{$\beta_U=0$} \\
		$\Pr(Y=1)$ & 0.04 & 0.00 & 0.04 & 0.00 & 0.04 & 0.00 & 0.04 & 0.00 & 0.04 & 0.00 \\ 
		Naive $\beta_1$ & 1.00 & 0.05 & 1.00 & 0.05 & 1.00 & 0.04 & 1.00 & 0.04 & 1.00 & 0.04 \\ 
		$\beta_2$ & 0.50 & 0.05 & 0.50 & 0.05 & 0.50 & 0.05 & 0.50 & 0.04 & 0.50 & 0.05 \\
\hline
	\end{tabular}
\end{table}

\begin{table}[ht]
	\scriptsize
	\centering
	\caption{Simulation results, logistic regression, Setting IV.A}
	\begin{tabular}{rrrrrrrrrrr}
		\hline
		& \multicolumn{2}{c}{$\gamma_U=0$} & \multicolumn{2}{c}{$\gamma_U=0.5$} &
		\multicolumn{2}{c}{$\gamma_U=1$} & \multicolumn{2}{c}{$\gamma_U=2$}  & \multicolumn{2}{c}{$\gamma_U=4$}\\ 
		& Mean  &SD & Mean & SD & Mean & SD & Mean & SD & Mean & SD \\ 
		\hline
		$R^2_{X|G_{IV},Z,U}$ & 0.33 & 0.01 & 0.43 & 0.01 & 0.60 & 0.01 & 0.82 & 0.00 & 0.94 & 0.00 \\ 
		$R^2_{X|G_{IV},Z}$ & 0.33 & 0.01 & 0.29 & 0.01 & 0.20 & 0.01 & 0.09 & 0.01 & 0.03 & 0.00 \\ 
		\multicolumn{11}{c}{$\beta_U=0$} \\ 
		$\Pr(Y=1)$ & 0.06 & 0.00 & 0.06 & 0.00 & 0.06 & 0.00 & 0.05 & 0.00 & 0.05 & 0.00 \\ 
		Naive $\beta_1$ & 1.00 & 0.06 & 1.00 & 0.06 & 1.00 & 0.06 & 1.00 & 0.06 & 1.00 & 0.05 \\ 
		$\beta_2$ & 0.50 & 0.05 & 0.50 & 0.05 & 0.50 & 0.05 & 0.50 & 0.05 & 0.50 & 0.05 \\
\hline
	\end{tabular}
\end{table}

\begin{table}[ht]
	\scriptsize
	\centering
	\caption{Simulation results, logistic regression, Setting III.B}
	\begin{tabular}{rrrrrrrrrrr}
		\hline
		& \multicolumn{2}{c}{$\gamma_U=0$} & \multicolumn{2}{c}{$\gamma_U=0.5$} &
		\multicolumn{2}{c}{$\gamma_U=1$} & \multicolumn{2}{c}{$\gamma_U=2$}  & \multicolumn{2}{c}{$\gamma_U=4$}\\ 
		& Mean  &SD & Mean & SD & Mean & SD & Mean & SD & Mean & SD \\ 
		\hline
		$R^2_{X|G_{IV},Z,U}$ & 0.33 & 0.01 & 0.43 & 0.01 & 0.60 & 0.01 & 0.82 & 0.00 & 0.94 & 0.00 \\ 
		$R^2_{X|G_{IV},Z}$ & 0.33 & 0.01 & 0.29 & 0.01 & 0.20 & 0.01 & 0.09 & 0.01 & 0.03 & 0.00 \\ 
		\multicolumn{11}{c}{$\beta_U=0$} \\
		$\Pr(Y=1)$ & 0.03 & 0.00 & 0.03 & 0.00 & 0.03 & 0.00 & 0.03 & 0.00 & 0.03 & 0.00 \\ 
		Naive $\beta_1$ & 1.00 & 0.04 & 1.00 & 0.04 & 1.00 & 0.04 & 1.00 & 0.04 & 1.00 & 0.04 \\ 
		$\beta_2$ & 0.50 & 0.04 & 0.50 & 0.04 & 0.50 & 0.04 & 0.50 & 0.04 & 0.50 & 0.04 \\
\hline
	\end{tabular}
\end{table}

\begin{table}[ht]
	\scriptsize
	\centering
	\caption{Simulation results, logistic regression, Setting IV.B}
	\begin{tabular}{rrrrrrrrrrr}
		\hline
		& \multicolumn{2}{c}{$\gamma_U=0$} & \multicolumn{2}{c}{$\gamma_U=0.5$} &
		\multicolumn{2}{c}{$\gamma_U=1$} & \multicolumn{2}{c}{$\gamma_U=2$}  & \multicolumn{2}{c}{$\gamma_U=4$}\\ 
		& Mean  &SD & Mean & SD & Mean & SD & Mean & SD & Mean & SD \\ 
		$R^2_{X|G_{IV},Z,U}$ & 0.33 & 0.01 & 0.43 & 0.01 & 0.60 & 0.01 & 0.82 & 0.00 & 0.94 & 0.00 \\ 
		$R^2_{X|G_{IV},Z}$ & 0.33 & 0.01 & 0.29 & 0.01 & 0.20 & 0.01 & 0.09 & 0.01 & 0.03 & 0.00 \\ 
		\hline
		\multicolumn{11}{c}{$\beta_U=0$} \\
		$\Pr(Y=1)$ & 0.03 & 0.00 & 0.03 & 0.00 & 0.03 & 0.00 & 0.03 & 0.00 & 0.03 & 0.00 \\ 
		Naive $\beta_1$ & 1.00 & 0.05 & 1.00 & 0.05 & 1.00 & 0.05 & 1.00 & 0.05 & 1.01 & 0.05 \\ 
		$\beta_2$ & 0.50 & 0.04 & 0.50 & 0.04 & 0.50 & 0.04 & 0.50 & 0.04 & 0.50 & 0.04 \\
\hline
	\end{tabular}
\end{table}

\clearpage
\begin{figure}[h!]
	\centering
	\includegraphics[width=16cm,height=15cm]{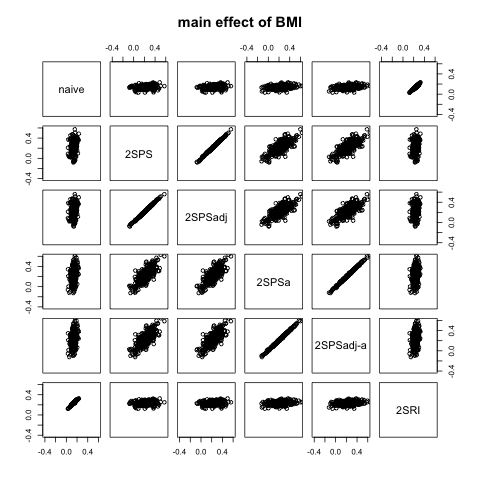}
	\caption{Pairwise plots of the estimates for the main effect of BMI for the naive estimators and the 5 MR estimators. }
\end{figure}

\clearpage
\begin{figure}[h!]
\centering
\includegraphics[width=16cm,height=15cm]{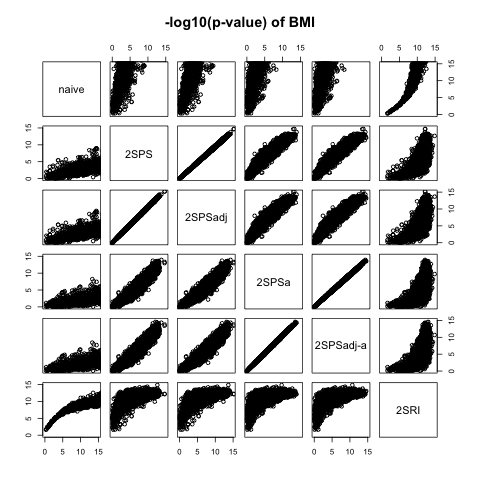}
\caption{Pairwise plots of $-\log_{10}\text{p-value}$ for the main effect of BMI for the naive estimators and the five MR-based estimators. }
\end{figure}

\clearpage
\begin{figure}[h!]
\centering
\includegraphics[width=16cm,height=15cm]{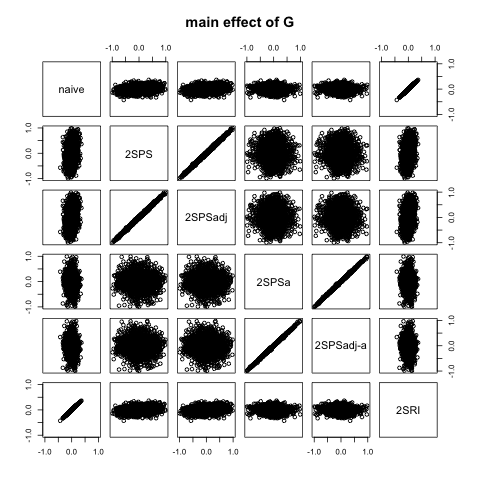}
\caption{Pairwise plots of the estimates for the main effect of G for the naive estimators and the five MR-based estimators. }
\end{figure}

\clearpage
\begin{figure}[h!]
\centering
\includegraphics[width=16cm,height=15cm]{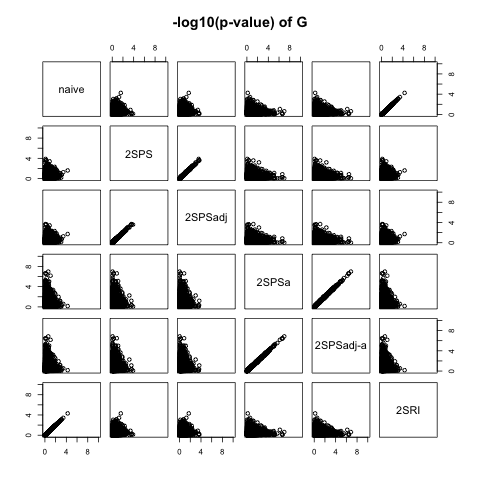}
\caption{Pairwise plots of 
	$-\log_{10}\text{p-value}$ for the main effect of G for the naive estimators and the five MR-based estimators. }
\end{figure}

\clearpage
\begin{figure}[h!]
\centering
\includegraphics[width=16cm,height=15cm]{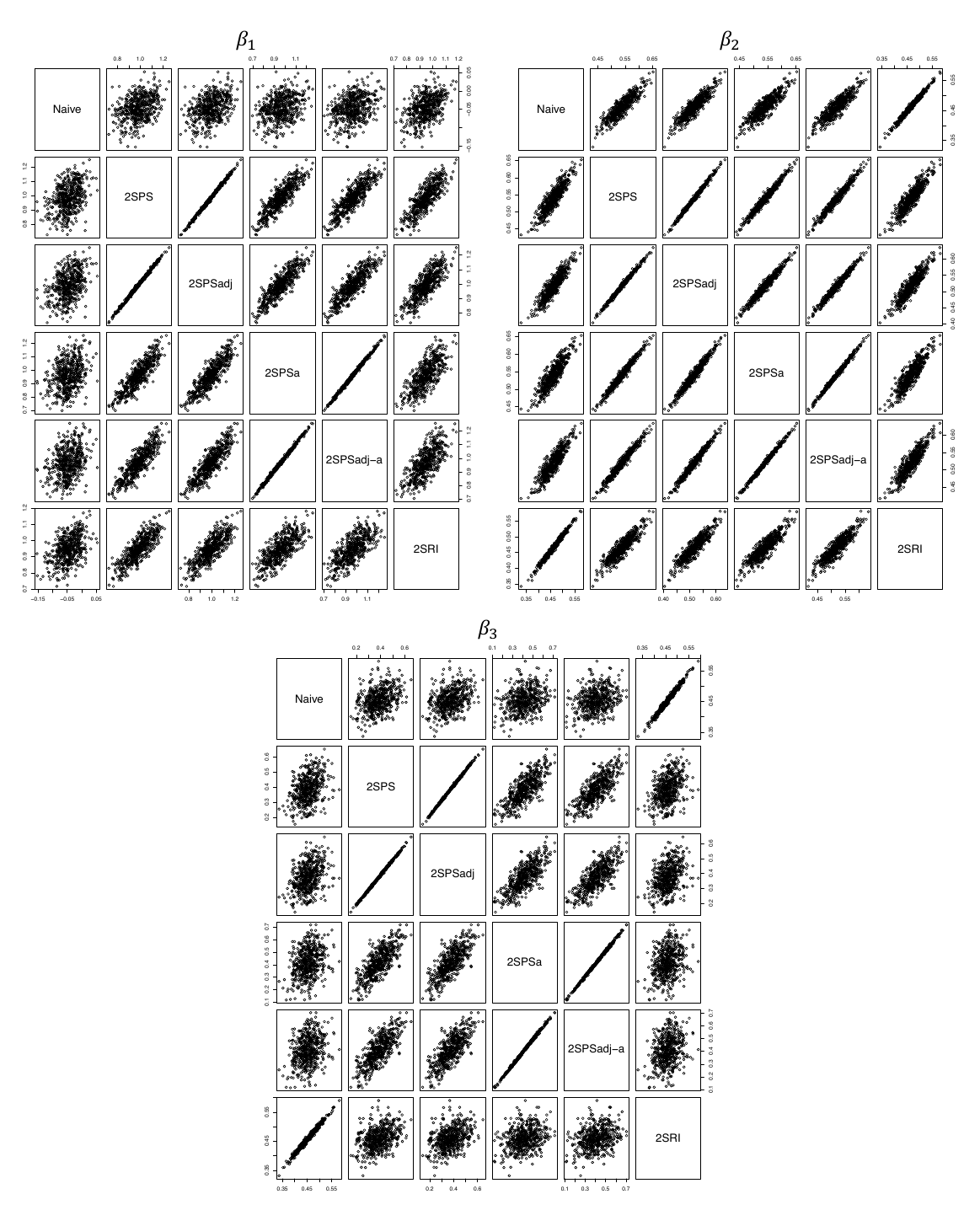}
\caption{Simulation results, logistic Regression, Setting III.A with $\gamma_U=-1$ and $\beta_U=1.5$: estimates pairwise plots}
\end{figure}

\clearpage

\noindent \textbf{FUNDING:}
Genetics and Epidemiology of Colorectal Cancer Consortium (GECCO): National Cancer Institute, National Institutes of Health, U.S. Department of Health and Human Services (U01 CA164930, R01 CA201407). Genotyping/Sequencing services were provided by the Center for Inherited Disease Research (CIDR) contract number HHSN268201200008I . This research was funded in part through the NIH/NCI Cancer Center Support Grant P30 CA015704. Scientific Computing Infrastructure at Fred Hutch funded by ORIP grant S10OD028685.

The ATBC Study is supported by the Intramural Research Program of the U.S. National Cancer Institute, National Institutes of Health, Department of Health and Human Services.

CLUE II funding was from the National Cancer Institute (U01 CA086308, Early Detection Research Network; P30 CA006973), National Institute on Aging (U01 AG018033), and the American Institute for Cancer Research. The content of this publication does not necessarily reflect the views or policies of the Department of Health and Human Services, nor does mention of trade names, commercial products, or organizations imply endorsement by the US government. Maryland Cancer Registry (MCR)  Cancer data was provided by the Maryland Cancer Registry, Center for Cancer Prevention and Control, Maryland Department of Health, with funding from the State of Maryland and the Maryland Cigarette Restitution Fund. The collection and availability of cancer registry data is also supported by the Cooperative Agreement NU58DP006333, funded by the Centers for Disease Control and Prevention. Its contents are solely the responsibility of the authors and do not necessarily represent the official views of the Centers for Disease Control and Prevention or the Department of Health and Human Services.

The Colon Cancer Family Registry (CCFR, www.coloncfr.org) is supported in part by funding from the National Cancer Institute (NCI), National Institutes of Health (NIH) (award U01 CA167551). Support for case ascertainment was provided in part from the Surveillance, Epidemiology, and End Results (SEER) Program and the following U.S. state cancer registries: AZ, CO, MN, NC, NH; and by the Victoria Cancer Registry (Australia) and Ontario Cancer Registry (Canada). The CCFR Set-1 (Illumina 1M/1M-Duo) and Set-2 (Illumina Omni1-Quad) scans were supported by NIH awards U01 CA122839 and R01 CA143237 (to GC). The CCFR Set-3 (Affymetrix Axiom CORECT Set array) was supported by NIH award U19 CA148107 and R01 CA81488 (to SBG). The CCFR Set-4 (Illumina OncoArray 600K SNP array) was supported by NIH award U19 CA148107 (to SBG) and by the Center for Inherited Disease Research (CIDR), which is funded by the NIH to the Johns Hopkins University, contract number HHSN268201200008I. Additional funding for the OFCCR/ARCTIC was through award GL201-043 from the Ontario Research Fund (to BWZ), award 112746 from the Canadian Institutes of Health Research (to TJH), through a Cancer Risk Evaluation (CaRE) Program grant from the Canadian Cancer Society (to SG), and through generous support from the Ontario Ministry of Research and Innovation. The SFCCR Illumina HumanCytoSNP array was supported in part through NCI/NIH awards U01/U24 CA074794 and R01 CA076366 (to PAN).  The content of this manuscript does not necessarily reflect the views or policies of the NCI, NIH or any of the collaborating centers in the Colon Cancer Family Registry (CCFR), nor does mention of trade names, commercial products, or organizations imply endorsement by the US Government, any cancer registry, or the CCFR.

COLO2\&3: National Institutes of Health (R01 CA060987).

Colorectal Cancer Transdisciplinary (CORECT) Study: The CORECT Study was supported by the National Cancer Institute, National Institutes of Health (NCI/NIH), U.S. Department of Health and Human Services (grant numbers U19 CA148107, R01 CA081488, P30 CA014089, R01 CA197350; P01 CA196569; R01 CA201407; R01 CA242218), National Institutes of Environmental Health Sciences, National Institutes of Health (grant number T32 ES013678) and a generous gift from Daniel and Maryann Fong.

CORSA:  The CORSA study was funded by Austrian Research Funding Agency (FFG) BRIDGE (grant 829675, to Andrea Gsur), the “Herzfelder’sche Familienstiftung” (grant to Andrea Gsur) and was supported by COST Action BM1206.

CPS-II: The American Cancer Society funds the creation, maintenance, and updating of the Cancer Prevention Study-II (CPS-II) cohort. The study protocol was approved by the institutional review boards of Emory University, and those of participating registries as required.

CRCGEN: Colorectal Cancer Genetics \& Genomics, Spanish study was supported by Instituto de Salud Carlos III, co-funded by FEDER funds –a way to build Europe– (grants PI14-613 and PI09-1286), Agency for Management of University and Research Grants (AGAUR) of the Catalan Government (grant 2017SGR723), Junta de Castilla y León (grant LE22A10-2), the Spanish Association Against Cancer (AECC) Scientific Foundation grant GCTRA18022MORE and the Consortium for Biomedical Research in Epidemiology and Public Health (CIBERESP), action Genrisk. Sample collection of this work was supported by the Xarxa de Bancs de Tumors de Catalunya sponsored by Pla Director d’Oncología de Catalunya (XBTC), Plataforma Biobancos PT13/0010/0013 and ICOBIOBANC, sponsored by the Catalan Institute of Oncology. We thank CERCA Programme, Generalitat de Catalunya for institutional support.

Czech Republic CCS: This work was supported by the Czech Science Foundation (21-04607X, 21-27902S), by the Grant Agency of the Ministry of Health of the Czech Republic (grants AZV NU21-07-00247 and AZV NU21-03-00145), and Charles University Research Fund (Cooperation 43-Surgical disciplines).

DACHS: This work was supported by the German Research Council (BR 1704/6-1, BR 1704/6-3, BR 1704/6-4, CH 117/1-1, HO 5117/2-1, HE 5998/2-1, KL 2354/3-1, RO 2270/8-1 and BR 1704/17-1), the Interdisciplinary Research Program of the National Center for Tumor Diseases (NCT), Germany, and the German Federal Ministry of Education and Research (01KH0404, 01ER0814, 01ER0815, 01ER1505A and 01ER1505B).

DALS: National Institutes of Health (R01 CA048998 to M. L. Slattery).

EDRN: This work is funded and supported by the NCI, EDRN Grant (U01-CA152753).

EPIC: The coordination of EPIC is financially supported by International Agency for Research on Cancer (IARC) and also by the Department of Epidemiology and Biostatistics, School of Public Health, Imperial College London which has additional infrastructure support provided by the NIHR Imperial Biomedical Research Centre (BRC). The national cohorts are supported by: Danish Cancer Society (Denmark); Ligue Contre le Cancer, Institut Gustave Roussy, Mutuelle Générale de l’Education Nationale, Institut National de la Santé et de la Recherche Médicale (INSERM) (France); German Cancer Aid, German Cancer Research Center (DKFZ), German Institute of Human Nutrition Potsdam- Rehbruecke (DIfE), Federal Ministry of Education and Research (BMBF) (Germany); Associazione Italiana per la Ricerca sul Cancro-AIRC-Italy, Compagnia di SanPaolo and National Research Council (Italy); Dutch Ministry of Public Health, Welfare and Sports (VWS), Netherlands Cancer Registry (NKR), LK Research Funds, Dutch Prevention Funds, Dutch ZON (Zorg Onderzoek Nederland), World Cancer Research Fund (WCRF), Statistics Netherlands (The Netherlands); Health Research Fund (FIS) - Instituto de Salud Carlos III (ISCIII), Regional Governments of Andalucía, Asturias, Basque Country, Murcia and Navarra, and the Catalan Institute of Oncology - ICO (Spain); Swedish Cancer Society, Swedish Research Council and and Region Skåne and Region Västerbotten (Sweden); Cancer Research UK (14136 to EPIC-Norfolk; C8221/A29017 to EPIC-Oxford), Medical Research Council (1000143 to EPIC-Norfolk; MR/M012190/1 to EPIC-Oxford). (United Kingdom).

ESTHER/VERDI. This work was supported by grants from the Baden-Württemberg Ministry of Science, Research and Arts and the German Cancer Aid.

Harvard cohorts: HPFS is supported by the National Institutes of Health (P01 CA055075, UM1 CA167552, U01 CA167552, R01 CA137178, R01 CA151993, and R35 CA197735), NHS by the National Institutes of Health (P01 CA087969, UM1 CA186107, R01 CA137178, R01 CA151993, and R35 CA197735), and PHS by the National Institutes of Health (R01 CA042182).

Hawaii Adenoma Study: NCI grants R01 CA072520.

Kentucky: This work was supported by the following grant support: Clinical Investigator Award from Damon Runyon Cancer Research Foundation (CI-8); NCI R01CA136726.

LCCS: The Leeds Colorectal Cancer Study was funded by the Food Standards Agency and Cancer Research UK Programme Award (C588/A19167).

MCCS cohort recruitment was funded by VicHealth and Cancer Council Victoria. The MCCS was further supported by Australian NHMRC grants 509348, 209057, 251553 and 504711 and by infrastructure provided by Cancer Council Victoria. Cases and their vital status were ascertained through the Victorian Cancer Registry (VCR) and the Australian Institute of Health and Welfare (AIHW), including the National Death Index and the Australian Cancer Database. BMLynch was supported by MCRF18005 from the Victorian Cancer Agency.

MEC: National Institutes of Health (R37 CA054281, P01 CA033619, and R01 CA063464).

MECC: This work was supported by the National Institutes of Health, U.S. Department of Health and Human Services (R01 CA081488, R01 CA197350, U19 CA148107, R01 CA242218, and a generous gift from Daniel and Maryann Fong.

NCCCS I \& II: We acknowledge funding support for this project from the National Institutes of Health, R01 CA066635 and P30 DK034987.

NFCCR: This work was supported by an Interdisciplinary Health Research Team award from the Canadian Institutes of Health Research (CRT 43821); the National Institutes of Health, U.S. Department of Health and Human Serivces (U01 CA074783); and National Cancer Institute of Canada grants (18223 and 18226). The authors wish to acknowledge the contribution of Alexandre Belisle and the genotyping team of the McGill University and Génome Québec Innovation Centre, Montréal, Canada, for genotyping the Sequenom panel in the NFCCR samples. Funding was provided to Michael O. Woods by the Canadian Cancer Society Research Institute.

PLCO: Intramural Research Program of the Division of Cancer Epidemiology and Genetics and supported by contracts from the Division of Cancer Prevention, National Cancer Institute, NIH, DHHS. Funding was provided by National Institutes of Health (NIH), Genes, Environment and Health Initiative (GEI) Z01 CP 010200, NIH U01 HG004446, and NIH GEI U01 HG 004438.

SMS and REACH: This work was supported by the National Cancer Institute (grant P01 CA074184 to J.D.P. and P.A.N., grants R01 CA097325, R03 CA153323, and K05 CA152715 to P.A.N., and the National Center for Advancing Translational Sciences at the National Institutes of Health (grant KL2 TR000421 to A.N.B.-H.)

The Swedish Low-risk Colorectal Cancer Study: The study was supported by grants from the Swedish research council; K2015-55X-22674-01-4, K2008-55X-20157-03-3, K2006-72X-20157-01-2 and the Stockholm County Council (ALF project).

Swedish Mammography Cohort and Cohort of Swedish Men: This work is supported by the Swedish Research Council /Infrastructure grant, the Swedish Cancer Foundation, and the Karolinska Institute´s Distinguished Professor Award to Alicja Wolk.

UK Biobank: This research has been conducted using the UK Biobank Resource under Application Number 8614

VITAL: National Institutes of Health (K05 CA154337).

WHI: The WHI program is funded by the National Heart, Lung, and Blood Institute, National Institutes
of Health, U.S. Department of Health and Human Services through contracts 75N92021D00001,
75N92021D00002, 75N92021D00003, 75N92021D00004, 75N92021D00005

\noindent \textbf{Acknowledgements:}

CCFR: The Colon CFR graciously thanks the generous contributions of their study participants, dedication of study staff, and the financial support from the U.S. National Cancer Institute, without which this important registry would not exist. The authors would like to thank the study participants and staff of the Seattle Colon Cancer Family Registry and the Hormones and Colon Cancer study (CORE Studies).

CLUE II: We thank the participants of Clue II and appreciate the continued efforts of the staff at the Johns Hopkins George W. Comstock Center for Public Health Research and Prevention in the conduct of the Clue II Cohort Study. Cancer data was provided by the Maryland Cancer Registry, Center for Cancer Prevention and Control, Maryland Department of Health, with funding from the State of Maryland and the Maryland Cigarette Restitution Fund. The collection and availability of cancer registry data is also supported by the Cooperative Agreement NU58DP006333, funded by the Centers for Disease Control and Prevention. Its contents are solely the responsibility of the authors and do not necessarily represent the official views of the Centers for Disease Control and Prevention or the Department of Health and Human Services.

CORSA: We kindly thank all individuals who agreed to participate in the CORSA study. Furthermore, we thank all cooperating physicians and students and the Biobank Graz of the Medical University of Graz.

CPS-II: The authors express sincere appreciation to all Cancer Prevention Study-II participants, and to each member of the study and biospecimen management group. The authors would like to acknowledge the contribution to this study from central cancer registries supported through the Centers for Disease Control and Prevention's National Program of Cancer Registries and cancer registries supported by the National Cancer Institute's Surveillance Epidemiology and End Results Program. The authors assume full responsibility for all analyses and interpretation of results. The views expressed here are those of the authors and do not necessarily represent the American Cancer Society or the American Cancer Society – Cancer Action Network.

Czech Republic CCS: We are thankful to all clinicians in major hospitals in the Czech Republic, without whom the study would not be practicable. We are also sincerely grateful to all patients participating in this study.

DACHS: We thank all participants and cooperating clinicians, and everyone who provided excellent technical assistance.

EDRN: We acknowledge all contributors to the development of the resource at University of Pittsburgh School of Medicine, Department of Gastroenterology, Department of Pathology, Hepatology and Nutrition and Biomedical Informatics.

EPIC: Where authors are identified as personnel of the International Agency for Research on Cancer/World Health Organization, the authors alone are responsible for the views expressed in this article and they do not necessarily represent the decisions, policy or views of the International Agency for Research on Cancer/World Health Organization.

Harvard cohorts: The study protocol was approved by the institutional review boards of the Brigham and Women’s Hospital and Harvard T.H. Chan School of Public Health, and those of participating registries as required. We acknowledge Channing Division of Network Medicine, Department of Medicine, Brigham and Women's Hospital as home of the NHS. The authors would like to acknowledge the contribution to this study from central cancer registries supported through the Centers for Disease Control and Prevention’s National Program of Cancer Registries (NPCR) and/or the National Cancer Institute’s Surveillance, Epidemiology, and End Results (SEER) Program.  Central registries may also be supported by state agencies, universities, and cancer centers.  Participating central cancer registries include the following: Alabama, Alaska, Arizona, Arkansas, California, Colorado, Connecticut, Delaware, Florida, Georgia, Hawaii, Idaho, Indiana, Iowa, Kentucky, Louisiana, Massachusetts, Maine, Maryland, Michigan, Mississippi, Montana, Nebraska, Nevada, New Hampshire, New Jersey, New Mexico, New York, North Carolina, North Dakota, Ohio, Oklahoma, Oregon, Pennsylvania, Puerto Rico, Rhode Island, Seattle SEER Registry, South Carolina, Tennessee, Texas, Utah, Virginia, West Virginia, Wyoming. The authors assume full responsibility for analyses and interpretation of these data.

Kentucky: We would like to acknowledge the staff at the Kentucky Cancer Registry.

LCCS: We acknowledge the contributions of Jennifer Barrett, Robin Waxman, Gillian Smith and Emma Northwood in conducting this study.

NCCCS I \& II: We would like to thank the study participants, and the NC Colorectal Cancer Study staff.

PLCO: The authors thank the PLCO Cancer Screening Trial screening center investigators and the staff from Information Management Services Inc and Westat Inc. Most importantly, we thank the study participants for their contributions that made this study possible. Cancer incidence data have been provided by the District of Columbia Cancer Registry, Georgia Cancer Registry, Hawaii Cancer Registry, Minnesota Cancer Surveillance System, Missouri Cancer Registry, Nevada Central Cancer Registry, Pennsylvania Cancer Registry, Texas Cancer Registry, Virginia Cancer Registry, and Wisconsin Cancer Reporting System. All are supported in part by funds from the Center for Disease Control and Prevention, National Program for Central Registries, local states or by the National Cancer Institute, Surveillance, Epidemiology, and End Results program. The results reported here and the conclusions derived are the sole responsibility of the authors.

SELECT: We thank the research and clinical staff at the sites that participated on SELECT study, without whom the trial would not have been successful. We are also grateful to the 35,533 dedicated men who participated in SELECT.

WHI: The authors thank the WHI investigators and staff for their dedication, and the study participants for making the program possible. A full listing of WHI investigators can be found at: http://www.whi.org/researchers/Documents

\end{document}